\documentclass[12pt, journal, onecolumn]{IEEEtran}
\usepackage{url}
\usepackage[dvips]{color}
\usepackage{algorithm}
\usepackage{algpseudocode}
\usepackage{multirow}
\usepackage{mdframed}
\usepackage{mathrsfs}
\usepackage{lipsum}
\usepackage{setspace}
\usepackage{graphicx}
\usepackage{graphics}
\usepackage{latexsym}
\usepackage{showidx}
\usepackage{latexsym}
\usepackage{amssymb}
\usepackage{amsfonts}
\usepackage{amsmath}
\usepackage{amssymb}
\usepackage{amsfonts}
\usepackage{amsmath}
\usepackage{bbold}
\usepackage{xcolor}
\usepackage[normalem]{ulem}
\newtheorem{theorem}{Theorem}

\newtheorem{definition}{Definition}
\newtheorem{problem}{Problem}

\newtheorem{proposition}{Proposition}

\usepackage{mathtools}
\DeclarePairedDelimiter\floor{\lfloor}{\rfloor}
\DeclarePairedDelimiter{\ceil}{\lceil}{\rceil}

\usepackage[tight,footnotesize]{subfigure}
\begin{document}
	\title{Hybrid-ARQ Based Relaying Strategies for Enhancing Reliability in Delay-Bounded Networks}
	\author{Jaya Goel and J. Harshan
	\thanks{Parts of this work have appeared in the 20th International Symposium on Modeling and Optimization in Mobile, Ad hoc, and Wireless Networks (WiOpt), in September 2022. The authors are with the Indian Institute of Technology Delhi, India.}}	
	\maketitle	
\vspace{-1cm}		
	\begin{abstract}
	Inspired by several delay-bounded mission-critical applications, this paper investigates chase-combining based hybrid automatic repeat request (CC-HARQ) protocols to achieve high reliability in delay-constrained applications. A salient feature of our approach is to use the end-to-end delay constraint for computing the total number of ARQs permitted in the network, and then optimally distributing them across the nodes so as to minimize packet-drop-probability (PDP), which is the end-to-end reliability metric of interest. Since the chase-combining strategy combines the received packets across multiple attempts, we observe that the PDP of the network depends on the coherence-time of the intermediate wireless channels. As a result, we address the question of computing optimal allocation of ARQs for CC-HARQ strategies under both slow-fading and fast-fading scenarios. For both the channel conditions, we derive closed-form expressions for the PDP, and then formulate several optimization problems for minimizing the PDP for a given delay-bound. Using extensive theoretical results on the local minima of the optimization problems, we synthesize low-complexity algorithms to obtain near-optimal ARQ distributions. Besides using extensive simulation results to validate our findings, a detailed end-to-end delay analysis is also presented to show that the proposed CC-HARQ strategies outperform already known Type-1 ARQ based strategies in several scenarios. 
	\end{abstract} 
	\maketitle	

	%
	\IEEEpeerreviewmaketitle	
	\section{Introduction}
	\label{sec:intro}
	In the fifth generation (5G) wireless networks and beyond, use-cases such as Internet-of-Things (IoT), massive Machine-type communications (mMTC) are pitched to play a vital role in deploying mission-critical applications  \cite{Schulz2017}. Examples for such mission-critical tasks include reliable and timely delivery of status updates in vehicular networks, facilitating real-time industrial automation tasks by deploying massive low-power IoT devices in factory settings, and so on. In these settings, one of the main challenges is to achieve highly reliable communication between a source and a destination with bounded delay constraints on the packets. Furthermore, generalizing the aforementioned mission-critical problem statements is also important in multi-hop wireless networks \cite{Badarneh2016, DavidChase1985}, which are known to facilitate expansion in the coverage area with power-limited wireless devices. Thus, in this work, we address the challenges in envisioning high reliability for delay-bounded applications over multi-hop wireless networks \cite{Shaikh2018, Ji2018}.  
	
As part of the ongoing research in wireless multi-hop networks, \cite{wiopt, our_work_TWC_1} recently proposed a retransmission based decode-and-forward (DF) based relaying strategy, wherein reliability within a hop is taken care by the number of Automatic Repeat Requests (ARQs) allotted to a given node, and the end-to-end delay on the packets is dictated by the total number of ARQs allotted to all the nodes in the multi-hop network. By using the information on the processing delays at each node, the total number of permitted retransmissions across the network, which is denoted by $q_{sum} \in \mathbb{Z}_{+}$, is estimated corresponding to the deadline on the packets. Subsequently, using $q_{sum}$ several questions are posed in \cite{our_work_TWC_1} on how to distribute these retransmissions across different nodes to maximize end-to-end reliability. In particular, the Type-1 ARQ protocol is employed, wherein the receiver node at each hop transmits an ACK or NACK depending on the success in decoding. Subsequently, on every retransmission, the receiver node discards the previous version of the packet and only uses the latest packet to decode the information. We note that Type-1 ARQs are suboptimal in slow-fading channels because there is no benefit in discarding the previous packets and then decoding the new packet (that is re-transmitted) under the same channel conditions. Also, in fast-fading channels, discarding the previous packets may result in more retransmissions thereby implying more delay. To address the mentioned issues, Hybrid Automatic Repeat Request (HARQ) is a promising technique wherein, on every retransmission, the receiver node combines the latest packet along with its previous copies to decode the information. Although an optimization problem on ARQ distribution has been studied in \cite{wiopt, our_work_TWC_1} for Type-1 ARQs, there is no such effort on HARQ to achieve high reliability in delay-bounded scenarios. Among many variants of HARQ schemes, a popular scheme is Chase-Combining HARQ (CC-HARQ), wherein each retransmission block is identical to the original code block, and all the received blocks are combined using the maximum-ratio combining  technique at the receiver \cite{Chaitanya_2013}.

	\subsection{Problem Statement}	
	We propose the CC-HARQ based DF strategies at each intermediate link for achieving high end-to-end reliability in multi-hop networks under delay-bounded scenarios. Towards that direction, it is clear that an appropriate number of ARQs must be allotted to intermediate links so that each link can utilize allotted ARQs by applying the CC-HARQ protocol. Furthermore, besides the Signal-to-Noise-Ratio (SNR) and the Line-of-Sight (LOS) components of the links, the end-to-end reliability of the network is dependent on (i) the intermediate channel conditions, for instance, whether the channel is slow-fading, wherein the channel realizations across the retransmissions are static, or fast-fading, wherein the channels across the retransmissions are statistically independent, and (ii) the specific protocol used to implement the CC-HARQ based DF strategies. Citing these dependencies, we address the following questions: (i) \emph{How to compute the optimal ARQ distribution at the intermediate links in order to achieve high-reliability for a given $q_{sum}$ in a CC-HARQ based multi-hop network?}, and (ii) \emph{How does imposing $q_{sum}$ total number of ARQs help in achieving the end-to-end delay of a multi-hop network?}
	
	\subsection{Contributions}
	\label{sec:contri}
We propose CC-HARQ based strategies for multi-hop networks with slow-fading channels (denoted as CC-HARQ-SF), wherein the channels are assumed to be static over the allotted attempts at each link. We also propose CC-HARQ based strategies for fast-fading scenarios (denoted as CC-HARQ-FF), wherein the channels are statistically independent across allotted attempts at each link. Under these scenarios, we propose two types of strategies, namely: non-cumulative strategy, wherein the intermediate nodes only have the knowledge of ARQs allotted to themselves but not to others, and the fully-cumulative strategy, wherein the intermediate nodes have the knowledge on ARQs allotted to the other nodes in addition to themselves. When the processing delay at each node is accurately estimated to compute $q_{sum}$, it is well known that the packets that violate the deadline are the ones that are dropped in the network \cite{our_work_TWC_1}. Therefore, for reliability analysis, we use the Packet-Drop-Probability (PDP) metric, which is defined as the fraction of packets that do not reach the destination. For the non-cumulative strategy, we derive a closed-form expression on PDP and formulate an optimization problem of minimizing the PDP for a given $q_{sum}$. We show that the optimization problem is non-tractable as it contains Marcum-Q functions. Towards obtaining near-optimal ARQ distributions, we propose a tight approximation on the first-order Marcum-Q functions, and then present non-trivial theoretical results for synthesizing a low-complexity algorithm. Through extensive simulations, we show that our analysis on the near-optimal ARQ distribution gives us the desired results with affordable complexity (see Section \ref{sec:analysis_SF} and Section \ref{sec:analysis_FF} for slow-fading and fast-fading, respectively). For the fully-cumulative strategy, we provide theoretical results on the optimal ARQ distribution in closed-form, and show that it provides lower PDP than that of the non-cumulative strategy (see Section \ref{sec:fully_cumm_SF} and see Section \ref{sec:fully_cumm_FF} for slow-fading and fast-fading, respectively). However, we also point out that the PDP benefits offered by the fully-cumulative strategy come at the cost of marginal increase in communication-overhead, as they make use of a counter in the packet to forward the residual ARQs to the rest of the nodes in the up-stream. Specifically, for a given $q_{sum}$, the communication-overhead is $\ceil{\log_{2}q_{sum}}$ bits, which is the maximum number of bits needed to represent the residual ARQs. Note that the communication-overhead is marginal when $\ceil{\log_{2}q_{sum}}$ is sufficiently small compared to the packet size.    
When the processing delay at each node is incorrectly estimated to compute $q_{sum}$, then a fraction of packets that reach the destination may violate the end-to-end deadline \cite{our_work_TWC_1}. Therefore, in such scenarios, we present a detailed analysis on end-to-end delay for each of our strategies (Sections \ref{sec:Sims_delay_analysis_SF} and \ref{sec:Sims_delay_analysis_FF}). 

\subsection{Related Work and Novelty}
\label{sec:rw}
The contribution in \cite{our_work_TWC_1} seems closest to the contributions in this work. However, as Type-2 ARQs are used in our work, the optimization problems and the approaches toward their solutions are different from \cite{our_work_TWC_1}, which were not dealt with bounded-delay applications hitherto. Therefore, this work cannot be considered as a straightforward extension of \cite{our_work_TWC_1}. There are other contributions in the literature \cite{Faisal_2021}-\cite{Tuninetti_2011}, \cite{Chaitanya_2013}, \cite{Shen_2011}, \cite{Stanojev_2008}, \cite{TVS_HARQ_UL} that use different forms of HARQ strategies. None of these contributions has addressed the computation of optimal ARQ distribution for the delay-bounded applications. Although  \cite{Stanojev_2008} appears similar to our contribution, it focusses on computing the optimal number of hops when the nodes implement the CC-HARQ strategy for a given deadline constraint. In contrast, we fix the number of hops and the deadline constraint, and then discuss how to distribute a certain number of retransmissions across the nodes. These two models are not equivalent because the way in which we introduce $q_{sum}$ is a function of the end-to-end deadline as well as the delay numbers introduced by the individual relays. However, \cite{Stanojev_2008} does not consider the delays introduced by the underlying relay nodes. A preliminary version of this work is available in \cite{our_work_WiOpt2022}, wherein the CC-HARQ-SF strategy is proposed for non-cumulative network. In addition to the contents of \cite{our_work_WiOpt2022}, this manuscript deals with the CC-HARQ-SF strategy for fully-cumulative network and the CC-HARQ-FF strategy for both non-cumulative and fully-cumulative networks.   
	\section{CC-HARQ Based Multi-Hop Network Model}
	\label{sec:system_model}
\begin{figure}[h!]
		\centering \includegraphics[scale = 0.93]{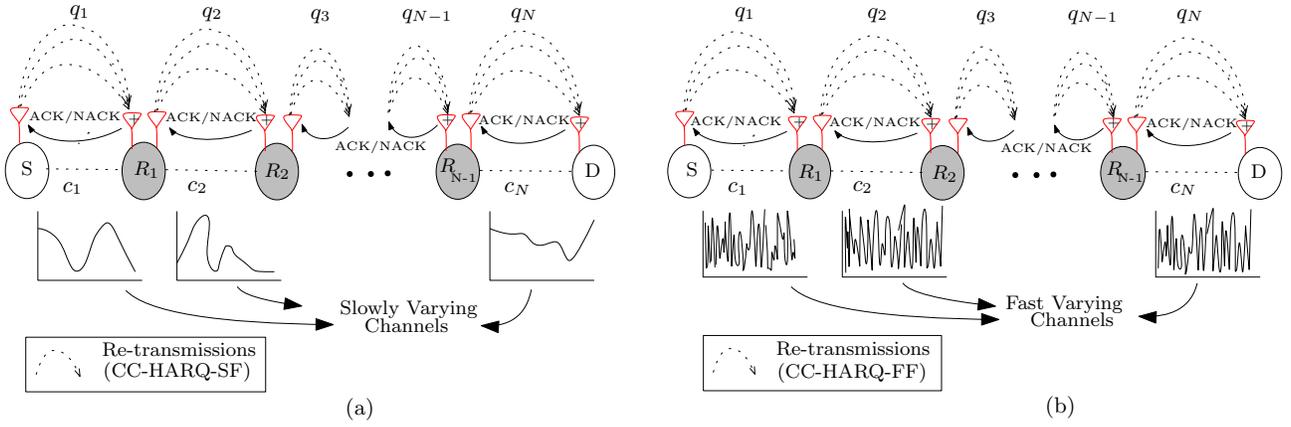}
		\vspace{-0.3cm}
		\centering{\caption{Illustration of an $N$-hop network with a source node $(S)$, the relay nodes $R_{1}, \ldots, R_{N-1}$ and the destination node $(D)$ following the CC-HARQ protocol at each intermediate link. Also, each intermediate link can be characterized by an LOS component $c_{k} \  \forall k \in [N]$ and the non-LOS component captures the slowly varying and fast varying behavior of the channel. 
				\label{fig:system_model}}}
	\end{figure}

	Consider a network with $N$-hops, as shown in Fig. \ref{fig:system_model}, consisting of a source node (S), a set of $N-1$ relays $R_{1}, R_{2}, \ldots, R_{N-1}$ and a destination node (D). By aggregating the information bits in the form of packets, we communicate these packets from S to D by using the $N-1$ intermediate relays. We assume that the channel between any two successive nodes is characterized by Rician fading by modelling the complex baseband channel of the $k$-th link, for $1 \leq k \leq N$, as $h_{k} = \sqrt{\frac{c_{k}}{2}}(1+\iota)+\sqrt{\frac{(1-c_{k})}{2}}g_{k},$ where $\iota = \sqrt{-1}$, $0 \leq c_{k} \leq 1$ is the LOS component, $1-c_{k}$ is the Non-LOS (NLOS) component, and $g_{k}$ is a Gaussian random variable with distribution $\mathcal{CN}(0,1)$. In this channel model, $c_{k}$ is a deterministic quantity, which characterizes different degrees of Rician fading channels, and also makes sure that $\mathbb{E}[|h_{k}|^{2}] = 1$ holds for any $c_{k}$. At the extreme ends, it is well known that $c_{k} = 0$ gives us the Rayleigh fading channels and $c_{k} = 1$ provides the Gaussian channels. 
	
	By assuming that the intermediate relays are sufficiently far apart from each other, throughout the paper, we use the vector $\mathbf{c}= [c_{1},c_{2},\ldots, c_{N}]$ to denote the LOS components of the $N$-hop network.
	Let $\mathcal{C} \subset \mathbb{C}^{L}$ denote the channel code used at the source of rate $R$ bits per channel use, i.e., $R = \frac{1}{L}\mbox{log}(|\mathcal{C}|).$ Let $\mathbf{x} \in \mathcal{C}$ denote a packet (which is a codeword in the code) transmitted by the source node such that its average energy per channel use is unity. When $\mathbf{x}$ is sent over the $k$-th link, for $1\leq k \leq N$, the corresponding baseband symbols gathered at the receiver of the $k$-th link over $L$ channel uses is $\mathbf{y}_{k} = h_{k}\mathbf{x} + \mathbf{w}_{k} \in \mathbb{C}^{L}$, where $\mathbf{w}_{k}$ is the additive white Gaussian noise (AWGN) at the receiver of the $k$-th link, distributed as $\mathcal{CN}(0,\sigma^{2}\mathbf{I}_{L})$. We assume that $h_{k}$ is known at the receiver of the $k$-th link owing to channel estimation; however, the transmitter of the $k$-th link does not know $h_{k}$. Since $h_{k}$ is sampled from an underlying distribution and the realization remains constant for $L$ channel uses, the transmission rate $R$ may not be less than the instantaneous mutual information of the $k$-th link. Therefore, in such cases, the corresponding relay node will fail to correctly decode the packet. The probability of such an event is
	\begin{eqnarray}
		\label{eq:outage_prob_link}
		P_{k} = \mbox{Pr} \Big( R > \log_{2}(1+ |h_{k}|^{2}\gamma ) \Big) = 1-Q_{1}\Bigg(\sqrt{\frac{2c_{k}}{(1-c_{k})}}, \sqrt{\frac{2(2^{R}-1)}{\gamma(1-c_{k})}}\Bigg),
	\end{eqnarray}
	where $\gamma =  \frac{1}{\sigma^{2}}$ is the average signal-to-noise-ratio (SNR) of the $k$-th link and $Q_{1}(\cdot,\cdot) $ is the first-order Marcum-Q function \cite{Marcum} obtained from the fact that $|h_{k}|^{2}$ is non-central chi-squared distributed with degrees of freedom two.\footnote{Note that this is a saddle-point approximation for the finite block-length error rates. Also, these approximations are tight when the block-length is of the order of a few hundreds \cite{V_Poor}.}
	To support the transmission rate, we follow the CC-HARQ strategy wherein a receiver node asks the transmitter node for retransmission of the packet and combines the received packet with the packets of the previous failed attempts to recover the information. In particular, for the given $N$-hop network model, a transmitter node gets an ACK or NACK from its successor node in the chain, indicating the success or failure of the transmission, respectively. Upon receiving a NACK, the transmitter retransmits the packet, and the receiver combines the current packet with the previously received copies of the packet. Let $q_{k}$ be the maximum number of attempts given to the transmitter of the $k$-th link to transmit the packet on demand. Consolidating the number of attempts given to each link, the ARQ distribution of the multi-hop network is represented by the vector $\mathbf{q}= [q_{1},q_{2},\ldots,q_{N}]$. Since we are addressing bounded-delay applications, we impose a sum constraint on the total number of retransmissions, given by $\sum_{i=1}^{N} q_{i} = q_{sum}$, for some $q_{sum} \in \mathbb{Z}_{+}$, which captures an upper bound on the end-to-end delay. This connection between the end-to-end deadline and $q_{sum}$ is established along the similar lines of the approach in \cite{our_work_TWC_1}, wherein the maximum number of permissible retransmissions is computed from the end-to-end deadline by assuming that the delay overheads from ACK/NACK in the reverse channel are sufficiently small compared to the payload. To understand how $q_{sum}$ captures an upper bound on end-to-end delay, suppose that the processing time at each hop is $\tau_{p}$ seconds (which includes packet encoding and decoding time), and the delay incurred for packet transmission at each hop is $\tau_{d}$ seconds (which includes the propagation delay and the time-frame of the packet), and the delay incurred because of NACK overhead is $\tau_{NACK}$ (which is the time taken for the transmitter to receive the NACK). Given the stochastic nature of the wireless channel at each link, the total number of packet retransmissions before the packet reaches the destination is a random variable, denoted by $n$, and as a result, the end-to-end delay between the source and the destination is upper bounded by $n \times(\tau_{p} + \tau_{d} + \tau_{NACK})$ seconds. In particular, when $\tau_{NACK} << \tau_{p} + \tau_{d}$, the end-to-end delay can be approximated as $n \times(\tau_{p} + \tau_{d})$ seconds. Thus, when the packet size and the decoding protocol at each node are established, and when the deadline on end-to-end delay (denoted by $\tau_{total})$ is known, we may impose an upper bound on $n$, provided by $q_{sum} = \floor{\frac{\tau_{total}}{\tau_{p}+\tau_{d}}}$. Here, we emphasize that $\tau_{total}$ is the real constraint, while $q_{sum}$ is only an intermediate parameter introduced for convenience. Note that when calculating $q_{sum}$ in the above discussion, queuing delays at the relay nodes have not been incorporated by assuming dedicated relays for a given source-destination pair. However, in a general setting, queuing delays can also be incorporated through the processing delays in a similar fashion.
	
When using the CC-HARQ protocol with the ARQ distribution $\mathbf{q}= [q_{1},q_{2},\ldots,q_{N}]$, the error event at the $k$-th link after using the $q_{k}$ attempts can be written as $P_{kq_{k}}= \mbox{Pr} \left(R > \log_{2}(1+ \sum_{j=1}^{q_{k}}|h_{kj}|^{2} \gamma)\right)$, where $h_{kj}$ is the channel realization at the $k$-th link for the $j$-th attempt. The expression on $P_{kq_{k}}$ is obtained due to the maximum ratio combining technique. For CC-HARQ-SF, i.e., when the channel realization of the $k$-th link remains static for the entire $q_{k}$ attempts, we denote $P_{kq_{k}}$ by $P_{kq_{k}}^{s}$, which can be re-written as $P_{kq_{k}}^{s} = \mbox{Pr} \left(R > \log_{2}(1+ |h_{k1}|^{2}q_{k} \gamma)\right)$. Since the channel remains constant over all the attempts, the channel realization in the first attempt, denoted by $h_{k1}$, is used in the expression for $P_{kq_{k}}^{s}$. Here, the superscript $`s'$ is used to represent the slow-fading scenario. Note that $P_{kq_{k}}^{s}$ is a function of first-order Marcum-Q function because $|h_{k1}|^{2}$ can be characterized by the non-central chi-square distribution with degrees of freedom $2$. It can be observed that on every retransmission, the packet gets added to its previous copies due to the CC-HARQ protocol, which results in an increased effective SNR of the given link. Similarly, for CC-HARQ-FF, wherein the channel realizations across the $q_{k}$ attempts are statistically independent, we denote $P_{kq_{k}}$ by $P_{kq_{k}}^{f}$, which can be written as $P_{kq_{k}}^{f}= \mbox{Pr} \left(R > \log_{2}(1+ \sum_{j=1}^{q_{k}}|h_{kj}|^{2} \gamma)\right)$. In this case, we highlight that only the NLOS component across the attempts take independent realizations, whereas the LOS components remain $c_{k}$. Here, the superscript $`f'$ is used to represent the fast-fading scenario across the attempts. Also, $P_{kq_{k}}^{f}$ is a function of $q_{k}$-th order Marcum-Q function because $\sum_{j=1}^{q_{k}}|h_{kj}|^{2}$ can be characterized by the non-central chi-square distribution with degrees of freedom $2q_{k}$. In the following sections, we introduce an end-to-end reliability metric for the CC-HARQ protocol under both slow-fading and fast-fading scenarios, and then discuss ``how to distribute a total of $q_{sum}$ ARQs across the nodes so as to maximize the end-to-end reliability metric?"
\section{Optimal ARQ distribution in CC-HARQ-SF for Non-cumulative Network}
	\label{sec:analysis_SF}
	In this section, we analyse the optimal ARQ distribution for the CC-HARQ-SF network, which is as given in Fig. \ref{fig:system_model}(a). We consider a protocol where each node knows the ARQs allotted to it and does not have the knowledge of the ARQs allotted to the other nodes. As a result, the transmitter of the $k$-th hop has a maximum of $q_{k}$ attempts to successfully forward the packet; otherwise, the packet is said to be dropped in that link. We refer the network with this protocol as the non-cumulative network. Since the packet can be dropped in any link, we use PDP as our end-to-end reliability metric of interest, given by 
	\begin{equation}
		\label{eq:pdp_expression_SF}
		p_{d}^{s} = P_{1q_{1}}^{s} + \sum_{k=2}^{N} P_{kq_{k}}^{s}\Bigg( \prod_{j=1}^{k-1}  (1-P_{jq_{j}}^{s})\Bigg), 
	\end{equation}
	where $P_{kq_{k}}^{s} = \mbox{Pr} \left(R > \log_{2}(1+ |h_{k1}|^{2}q_{k} \gamma)\right)$. Now, we are interested in allocating the ARQs to the nodes to minimize the PDP in \eqref{eq:pdp_expression_SF} for a given sum constraint $\sum_{i = 1}^{N} q_{i} = q_{sum}$. Henceforth, we formulate our optimization problem in Problem \ref{opt_problem_SF} as shown below. Throughout this paper, we refer to the solution of Problem \ref{opt_problem_SF} as the optimal ARQ distribution for CC-HARQ-SF based non-cumulative network. 
	\begin{mdframed}
		\begin{problem}
			\label{opt_problem_SF}
			For a given $\mathbf{c}$, $\gamma = \frac{1}{\sigma^{2}}$ and $q_{sum}$, solve $q_{1}^{*},q_{2}^{*},\ldots q_{N}^{*}=\arg\underset{q_{1},q_{2},\ldots q_{N}}{\text{min}}\ p_{d}^{s},$ 
				\text{subject to} $\ q_{k} \geq 1, q_{k}\in \mathbb{Z_{+}} \ \forall k \in [N],
				\sum_{i=1}^{N}q_{i} = \ q_{sum}.$ 		
		\end{problem}
	\end{mdframed}
	As the PDP expression in \eqref{eq:pdp_expression_SF} is dependent on the first-order Marcum-Q function, tackling the Marcum-Q function analytically is challenging because it contains a modified Bessel function of the first kind \cite{Marcum}. Therefore, towards solving Problem \ref{opt_problem_SF}, we first present an approximation of the Marcum-Q function at high SNR and use it to obtain the necessary and sufficient conditions on the near-optimal ARQ distribution. 
	
	\subsection{Approximation on the First-Order Marcum-Q Function}
	\label{approximation_1}
	\begin{theorem}
		\label{thm:approximation_Marcum_Q_1}
		For a given CC-HARQ-SF based $N$-hop network, at a high SNR regime, we can write the first-order Marcum-Q function, denoted by $Q_{1}(a_{k}, b_{k})$, as $Q_{1}(a_{k},b_{k}) =  \tilde{Q}_{1}(a_{k},b_{k}) + O(b_{k}^{4})$, where $\tilde{Q}_{1}(a_{k},b_{k}) \triangleq 1 - \frac{b_{k}^{2}}{2} e^{\frac{-a_{k}^{2}}{2}}$ for all $k \in [N]$, and $a_{k}$ and $b_{k}$ are related to the SNR, $c_{k}$ and $R$. Furthermore, at high SNR, we have $Q_{1}(a_{k},b_{k}) \approx  \tilde{Q}_{1}(a_{k},b_{k})$.
	\end{theorem}
	\begin{IEEEproof} 
   In the context of the $N$-hop network, as shown in Fig. \ref{fig:system_model} (a) (in Section \ref{sec:system_model}), the Marcum-Q function of first-order associated with the $k$-th hop is given by 
\begin{eqnarray*}
	Q_{1}(a_{k},b_{k}) = e^{-\frac{a_{k}^{2}}{2}}e^{-\frac{b_{k}^{2}}{2}}\sum_{i=0}^{\infty}\bigg(\frac{a_{k}^{2}}{2}\bigg)^{i} \bigg(\sum_{m=0}^{\infty} \bigg(\frac{a_{k}b_{k}}{2}\bigg)^{2m}  \frac{1}{m!(m+i)!}\bigg),  
\end{eqnarray*}
where $a_{k}= \sqrt{(2c_{k})/(1-c_{k})}$, $b_{k}= \sqrt{2(2^{R}-1)/\gamma(1-c_{k})}$ such that $\gamma = 1/\sigma^{2}$. We highlight that $a_{k}$ depends on the LOS component and $b_{k}$ is dependent on SNR. After expanding the above equation, we can write $Q_{1}(a_{k},b_{k}) = \tilde{Q}_{1}(a_{k},b_{k}) + \epsilon_{Q_{1}}$, where $\epsilon_{Q_{1}}$ is given by 
\begin{small}
			\begin{eqnarray*}
			\epsilon_{Q_{1}}&=& \frac{b_{k}^{4}}{4}(e^{-a_{k}^{2}/2}- 1) + e^{-a_{k}^{2}/2}\Bigg(\sum_{j=2}^{\infty} \frac{(-b_{k}^{2}/2)^{j}}{j!}\Bigg)\Bigg[\sum_{i=0}^{\infty}\Bigg(\frac{a_{k}^{2}}{2}\Bigg)^{i} \sum_{m=2}^{\infty}\Bigg(\frac{a_{k}b_{k}}{2}\Bigg)^{2m}   \frac{1}{m!(m+i)!}\Bigg] + \\ & &  e^{-a_{k}^{2}/2}
				\Bigg(\sum_{i=0}^{\infty}\frac{(-b_{k}^{2}/2)^{i}}{i!}\Bigg)\Bigg[\sum_{i=0}^{\infty}\Bigg(\frac{a_{k}^{2}}{2}\Bigg)^{i} \sum_{m=2}^{\infty}\Bigg(\frac{a_{k}b_{k}}{2}\Bigg)^{2m} \frac{1}{m!(m+i)!}\Bigg].
			\end{eqnarray*}  
		\end{small} 
\noindent Since only $b_{k}$ is dependent on the SNR, we focus on the structure of $\tilde{Q}_{1}(a_{k},b_{k})$ and $\epsilon_{Q_{1}}$ with respect to $b_{k}$. Evidently, $\epsilon_{Q_{1}}$ contains terms of the form $(b_{k}^{2}/2)^{p}$, for $p \in \{2,3, \ldots, \infty\}$, whereas $\tilde{Q}_{1}(a_{k},b_{k})$ contains $b_{k}^{2}/2$. Note that at high SNR, $b_{k}$ is very small, and therefore, $b_{k}^{4}$ dominates $b_{k}^{p}$, for $p \geq 6$, in $\epsilon_{Q_{1}}$. Thus, we can write $Q_{1}(a_{k},b_{k}) =  \tilde{Q}_{1}(a_{k},b_{k}) + O(b_{k}^{4})$ at high SNR values. Finally, along the similar lines, we obtain $Q_{1}(a_{k},b_{k}) \approx \tilde{Q}_{1}(a_{k},b_{k}) \triangleq 1 - \frac{b_{k}^{2}}{2} e^{\frac{-a_{k}^{2}}{2}},$ by neglecting the higher powers of $b_{k}$. For more details on this proof, we refer the reader to \cite[Theorem 1]{our_work_WiOpt2022}.
	\end{IEEEproof}
	To validate the accuracy of the approximation for different values of SNR and the LOS components, one can plot the original expression of the Marcum-Q function and its approximated expression, as shown in the preliminary version of this work in \cite{our_work_WiOpt2022}. Using the approximation on the Marcum-Q function in Theorem \ref{thm:approximation_Marcum_Q_1}, we can approximate $P_{kq_{k}}^{s}$ as $\tilde{P}_{kq_{k}}^{s}$, given by
	\begin{equation}
		\label{eq:approximation_P_k}
		\tilde{P}^{s}_{kq_{k}} \triangleq \frac{\phi}{q_{k}(1-c_{k})} e^{\frac{-a_{k}^{2}}{2}},
	\end{equation}
	where $\phi = \frac{2^{R}-1}{\gamma}$. Finally, by using \eqref{eq:approximation_P_k}, we can write the approximate version of \eqref{eq:pdp_expression_SF} as
	\begin{equation}
		\label{eq:Approx_pdp_expression}
		\tilde{p}^{s}_{d} = \tilde{P}^{s}_{1q_{1}} + \sum_{k=2}^{N} \tilde{P}^{s}_{kq_{k}} \Bigg( \prod_{j=1}^{k-1}  (1-\tilde{P}^{s}_{jq_{j}})\Bigg),
	\end{equation} 
	where $\tilde{p}_{d}^{s}$ denotes the approximation on $p_{d}^{s}$. Henceforth, using the approximated expression on $p_{d}^{s}$ as above, we formulate an  optimization problem in Problem \ref{opt_problem_2} as shown below. Throughout this paper, we refer to the solution of Problem \ref{opt_problem_2} as the near-optimal ARQ distribution since the objective function in Problem \ref{opt_problem_2} is an approximation of the objective function in Problem \ref{opt_problem_SF}. In the next section, we present some properties on $\tilde{p}_{d}^{s}$ to solve Problem \ref{opt_problem_2}. 
	\begin{mdframed}
		\begin{problem}
			\label{opt_problem_2}
			For a given $\mathbf{c}$, a high SNR $\gamma = \frac{1}{\sigma^{2}}$ and $q_{sum}$, solve $q_{1}^{*},q_{2}^{*},\ldots q_{N}^{*}=\arg\underset{q_{1},q_{2},\ldots q_{N}}{\text{min}}\ \tilde{p}_{d}^{s}$
				\text{subject to} $q_{k} \geq 1, q_{k}\in \mathbb{Z_{+}} \ \forall k \in [N],
				\sum_{i=1}^{N}q_{i} = \ q_{sum}$.  
		\end{problem}
	\end{mdframed}	
	\subsection{Analysis on the Near-Optimal ARQ Distribution in CC-HARQ-SF Based Non-Cumulative Network}
	\label{sec:neccAndSuffConditions}
	Before we obtain the necessary and sufficient conditions on the optimal ARQ distribution, we show that a link with a higher LOS must not be given more ARQs than the link with a lower LOS.
	\begin{theorem}
		\label{LOShighlow_SF}
		For a given LOS vector $\mathbf{c}$ and a high SNR regime in a CC-HARQ-SF based multi-hop network, the solution to Problem \ref{opt_problem_2} satisfies the property that whenever $c_{i} \geq c_{j}$, we have $q_{i} \leq q_{j} \ \forall i,j \in [N]$. 
	\end{theorem}
	\begin{IEEEproof}
		To highlight the location of $c_{i}$ and $c_{j}$, we write $\mathbf{c}$ as  $[c_{1}, c_{2}, \ldots, c_{i}, \ldots, c_{j}, \ldots, c_{N-1}, c_{N}]$ such that $j > i$. Assume that $c_{j} > c_{i}$, and $q_{i}$ and $q_{j}$ denote the number of ARQs allotted to the $i$-th and $j$-th link, respectively. Furthermore, let us suppose that $q_{i} =q_{j} = q$ and we have one ARQ with us. Now, the problem is deciding whether to allocate that additional ARQ to the $i$-th link or the $j$-link that results in lower PDP. Towards solving this problem, we consider an equivalent multi-hop network with LOS vector $\mathbf{c}' = [c_{1}, c_{2}, \ldots, c_{N-1}, \ldots, c_{N}, \ldots, c_{i}, c_{j}]$, wherein $\mathbf{c}'$ is obtained from $\mathbf{c}$ by swapping $c_{i}$ with $c_{N-1}$ and $c_{j}$ with $c_{N}$. By using a similar approach given in \cite[Theorem 1]{our_work_TWC_1}, we can show that the PDP of the multi-hop networks with the LOS vectors $\mathbf{c}$ and $\mathbf{c}'$ are identical. Furthermore, the PDP of the $N$-hop network with LOS vector $\mathbf{c}'$, is written as 
		\begin{eqnarray*}
			\tilde{p}^{s}_{d}  =  \tilde{P}^{s}_{1q_{1}} + \tilde{P}^{s}_{2q_{2}}(1-\tilde{P}^{s}_{1q_{1}}) + \ldots + \left(\tilde{P}^{s}_{iq_{i}} + \tilde{P}^{s}_{jq_{j}}(1 -\tilde{P}^{s}_{iq_{i}})\right)\prod_{k \in [N] \setminus \{i, j\}} (1-\tilde{P}^{s}_{kq_{k}}). 
		\end{eqnarray*} 
		Note that $\tilde{P}^{s}_{iq_{i}}$ and $\tilde{P}^{s}_{jq_{j}}$ appear only in the last term of the above expression. Since the question of allocating the additional ARQ is dependent only on the expression $\tilde{P}^{s}_{iq_{i}} + \tilde{P}^{s}_{jq_{j}}(1 - \tilde{P}^{s}_{iq_{i}})$, we henceforth do not use the entire expression for PDP. Additionally, since $q_{i} = q_{j} = q$, 
		we obtain one of the following expressions when allocating the additional ARQ,  
		\begin{eqnarray*}
			A & =& \tilde{P}^{s}_{i(q+1)}+\tilde{P}^{s}_{jq}(1-\tilde{P}^{s}_{i(q+1)}),\\
			B & =& \tilde{P}^{s}_{iq} + \tilde{P}^{s}_{j(q+1)}(1-\tilde{P}^{s}_{iq}).
		\end{eqnarray*}
		Since $c_{i} < c_{j}$, we know that $\tilde{P}^{s}_{iq} > \tilde{P}^{s}_{jq}$ for $q=1$. To prove the statement of the theorem, we use the approximation \eqref{eq:approximation_P_k} to show that $A < B$. Therefore, using the approximation \eqref{eq:approximation_P_k}, we write 
		\begin{eqnarray*}
			\begin{aligned}
				A = \frac{\phi e^{\frac{-a_{i}^{2}}{2}}}{(q+1)(1-c_{i})}  + \frac{\phi e^{\frac{-a_{j}^{2}}{2}}}{q(1-c_{j})}\bigg(1-  \frac{\phi e^{\frac{-a_{i}^{2}}{2}}}{(q+1)(1-c_{i})}\bigg),\\
				B  = \frac{\phi e^{\frac{-a_{i}^{2}}{2}}}{q(1-c_{i})}  + \frac{\phi e^{\frac{-a_{j}^{2}}{2}}}{(q+1)(1-c_{j})}\bigg(1-  \frac{\phi e^{\frac{-a_{i}^{2}}{2}}}{q(1-c_{i})}\bigg).	
			\end{aligned}
		\end{eqnarray*}
		On solving the difference $A-B$, we get 
		\begin{eqnarray*}
			\phi\bigg( \dfrac{e^{-a_{j}^{2}/2}-e^{-a_{i}^{2}/2}+ c_{j}e^{-a_{i}^{2}/2}-c_{i}e^{-a_{j}^{2}/2}}{{(1-c_{i})(1-c_{j})}}\bigg ) \underbrace{\bigg(\dfrac{1}{q}-\dfrac{1}{q+1}\bigg)}. 
		\end{eqnarray*}
		In the above equation, note that the second product term in the bracket is positive. Therefore, to prove that $A-B<0$, first, we show that $e^{-a_{j}^{2}/2}-e^{-a_{i}^{2}/2}+ c_{j}e^{-a_{i}^{2}/2}-c_{i}e^{-a_{j}^{2}/2} \leq 0$ for any $i,j \in [N]$. To proceed further, we start by assuming that the above inequality is true, and by rewriting the above equation, we get $e^{-a_{j}^{2}/2}(1-c_{i})  \leq e^{-a_{i}^{2}/2}(1-c_{j})$ which in turns equal to $\dfrac{(1-c_{i})}{(1-c_{j})} \leq \dfrac{e^{-a_{i}^{2}/2}}{e^{-a_{j}^{2}/2}}$. Now, by taking the logarithm on both sides and on expanding both $a_{i}$ and $a_{j}$, we obtain the inequality $\log\bigg(\dfrac{1-c_{i}}{1-c_{j}}\bigg) \leq \dfrac{c_{j}-c_{i}}{(1-c_{i})(1-c_{j})}$. Let $x =\dfrac{1-c_{i}}{1-c_{j}} $, and therefore, $x-1 = \dfrac{1-c_{i}}{1-c_{j}} - 1 = \dfrac{c_{j}-c_{i}}{(1-c_{j})}$. Furthermore, by using $x$, we can rewrite the above inequality as $\log x \leq \dfrac{x-1}{(1-c_{i})}$, where $\dfrac{1}{1-c_{i}} \geq 1$ because $c_{i} \leq 1$. Moreover, by using the standard inequality i.e. $\log x \leq (x-1)$, we can prove that the inequality $\log x \leq \dfrac{x-1}{(1-c_{i})}$ is true for any $i, j \in [N]$. Hence, this completes the proof that $A-B<0$. 
	\end{IEEEproof}
		
	Henceforth, we use the set $\mathbb{S} = \{\mathbf{q} \in \mathbb{Z}_{+}^{N} \ | \ \sum_{j=1}^{N} q_{j} = q_{sum} \ \& \ q_{j} \geq 1 \ \forall j \}$ to define the search space for the optimal ARQ distribution. Furthermore, for a given $\mathbf{q} \in \mathbb{S}$, its neighbors are defined as below.
	\begin{definition}
		\label{def:nn}
		The set of neighbors for a given $\mathbf{q} \in \mathbb{S}$ is defined as $\mathcal{H}(\mathbf{q}) = \{\bar{\mathbf{q}} \in \mathbb{S} ~|~ d(\mathbf{q},\bar{\mathbf{q}}) = 2\},$ where $d(\mathbf{q},\bar{\mathbf{q}})$ denotes the Hamming distance between $\mathbf{q}$ and $\bar{\mathbf{q}}$.
	\end{definition}
	In the following definition, we present a local minima of the space $\mathbb{S}$ by evaluating the PDP of the CC-HARQ-SF based non-cumulative network over the vectors in $\mathbb{S}$.  
	\begin{definition}
		\label{def:lm}  
		To be a local minima of $\mathbb{S}$ for the CC-HARQ-SF strategy with the objective function $\tilde{p}_{d}^{s}$, an ARQ distribution $\mathbf{q}^{*} \in \mathbb{S}$ must satisfy the condition $\tilde{p}_{d}^{s}(\mathbf{q}^{*}) \leq \tilde{p}_{d}^{s}(\mathbf{q}) \mbox{~for every~} \mathbf{q} \in \mathcal{H}(\mathbf{q}^{*})$, such that $\tilde{p}_{d}^{s}(\mathbf{q}^{*})$ and $\tilde{p}_{d}^{s}(\mathbf{q})$ represent the PDP evaluated at the distributions $\mathbf{q}^{*}$ and $\mathbf{q}$, respectively.   
	\end{definition}
	Using the above definition, we derive a set of necessary and sufficient conditions on the local minima in the following theorem.
	\begin{theorem}
		\label{NeccSuff}
		For a given $N$-hop network with LOS vector $\mathbf{c}$, the ARQ distribution $\mathbf{q}^{*} = [q^{*}_{1}, q^{*}_{2}, \ldots, q^{*}_{N}]$ is said to be a local minima for CC-HARQ-SF strategy if and only if $q^{*}_{i}$ and $q^{*}_{j}$ for $i \neq j$ satisfy the following bounds
		 \begin{eqnarray}
			\label{LBUB_1}
			q^{*2}_{j}K_{i}- q^{*}_{j}(K_{i}+K_{j}K_{i}) -C_{1} & \leq & 0  ,\\
			\label{LBUB_2}
			q^{*2}_{j}K_{i}+ q^{*}_{j}(K_{i}-K_{j}K_{i})-C_{2} & \geq & 0,
		\end{eqnarray} 
		\noindent where $C_{1} = -K_{j}K_{i} + q^{*2}_{i}K_{j} +  q^{*}_{i}(K_{j}-K_{j}K_{i}) $, $C_{2} = K_{j}K_{i} + q^{*2}_{i}K_{j}-  q^{*}_{i}(K_{j}+K_{j}K_{i}) $ with $K_{t}$ for $t \in \{i,j\}$ given by
		\begin{equation}
		\label{eq_kt}
		K_{t} =\dfrac{\phi}{1-c_{t}}e^{\bigg(\dfrac{-c_{t}}{1-c_{t}}\bigg)}, \mbox{ where} \ \phi = \frac{2^{R}-1}{\gamma}.
		\end{equation}
	\end{theorem}	
	\begin{IEEEproof}
		According to Definition \ref{def:nn}, it can be observed that a neighbor of $\mathbf{q}^{*}$ in the search space $\mathbb{S}$ differs in two positions with respect to $\mathbf{q}^{*}$. Let these neighbors are of the form $\hat{\mathbf{q}}_{+} = [q^{*}_{1}, q^{*}_{2}, \ldots, q^{*}_{i} + 1, \ldots, q^{*}_{j} -1, \ldots, q^{*}_{N}]$ and $\hat{\mathbf{q}}_{-} = [q^{*}_{1}, q^{*}_{2}, \ldots, q^{*}_{i} - 1, \ldots, q^{*}_{j} + 1, \ldots, q^{*}_{N}]$ that differs in two positions at $i$ and $j$ provided $q^{*}_{i} - 1 \geq 1$ and $q^{*}_{j} - 1 \geq 1$. Because of the type of search space and the expression of PDP, we invoke the results from \cite[Theorem 1]{our_work_TWC_1} that the PDP remains identical after swapping intermediate links. Therefore, instead of considering the multi-hop network with LOS vector $\mathbf{c} = [c_{1}, c_{2}, \ldots, c_{i}, \ldots, c_{j}, \ldots, c_{N-1}, c_{N}]$, we consider its permuted version with the LOS vector $\mathbf{c} = [c_{1}, c_{2}, \ldots, c_{N-1}, \ldots, c_{N}, \ldots, c_{i}, c_{j}]$, wherein the $i$-th link is swapped with $(N-1)$-th link, and the $j$-th link is swapped with $N$-th link. Consequently, the local minima and its two neighbors are respectively of the form $\mathbf{q}^{*} = [q^{*}_{1}, q^{*}_{2}, \ldots, q^{*}_{N-1}, \ldots, q^{*}_{N}, \ldots, q^{*}_{i}, q^{*}_{j}]$, $\hat{\mathbf{q}}_{+} = [q^{*}_{1}, q^{*}_{2}, \ldots, q^{*}_{N-1}, \ldots, q^{*}_{N}, \ldots, q^{*}_{i} + 1, q^{*}_{j} - 1]$ and $\hat{\mathbf{q}}_{-} = [q^{*}_{1}, q^{*}_{2}, \ldots, q^{*}_{N-1}, \ldots, q^{*}_{N}, \ldots, q^{*}_{i} - 1, q^{*}_{j} + 1]$. By using the definition of local minima, we have the inequalities
	\begin{eqnarray*}
		\label{eqr1}
		\tilde{p}^{s}_{d}(\mathbf{q}^{*}) \leq \tilde{p}^{s}_{d}(\hat{\mathbf{q}}_{+}), \mbox{ and } \tilde{p}^{s}_{d}(\mathbf{q}^{*}) \leq \tilde{p}^{s}_{d}(\hat{\mathbf{q}}_{-}),
	\end{eqnarray*}
	where $\tilde{p}^{s}_{d}(\mathbf{q}^{*})$, $\tilde{p}^{s}_{d}(\hat{\mathbf{q}}_{+})$ and $\tilde{p}^{s}_{d}(\hat{\mathbf{q}}_{-})$ represent the PDP evaluated at the distributions $\mathbf{q}^{*}$, $\hat{\mathbf{q}}_{+}$, and $\hat{\mathbf{q}}_{-}$, respectively. Because of the fact that $\hat{\mathbf{q}}_{+}$ and $\hat{\mathbf{q}}_{-}$ differ only in the last two positions and the structure of the PDP, it is possible to demonstrate that the above inequalities are equivalent to 
	\begin{eqnarray}
		\label{express1}
		\tilde{P}^{s}_{iq^{*}_{i}}+ \tilde{P}^{s}_{jq^{*}_{j}}\left(1-\tilde{P}^{s}_{iq^{*}_{i}}\right) \leq  \tilde{P}^{s}_{i(q^{*}_{i}+1)} + \tilde{P}^{s}_{j(q^{*}_{j}-1)} (1- \tilde{P}^{s}_{i(q^{*}_{i}+1)}),\\
		\label{express2}
		\tilde{P}^{s}_{iq^{*}_{i}}+\tilde{P}^{s}_{jq^{*}_{j}}\left(1-\tilde{P}^{s}_{iq^{*}_{j}}\right) \leq \tilde{P}^{s}_{i(q^{*}_{i}-1)} + \tilde{P}^{s}_{j(q^{*}_{j}+1)}(1- \tilde{P}^{s}_{i(q^{*}_{i}-1)}),	
	\end{eqnarray} respectively. 
	First, let us proceed with the \eqref{express1} to derive a necessary and sufficient conditions on $q^{*}_{i}$ and $q^{*}_{j}$. On expanding $\tilde{P}^{s}_{iq}$ and $\tilde{P}^{s}_{jq}$, and by using the approximation \eqref{eq:approximation_P_k} on outage probability, \eqref{express1} can be rewritten as 
	\begin{equation*}
		\dfrac{K_{i}}{q_{i}^{*}} + 	\dfrac{K_{j}}{q_{j}^{*}}\bigg(1-\dfrac{K_{i}}{q_{i}^{*}}\bigg) \leq \dfrac{K_{i}}{q_{i}^{*}+1} + 	\dfrac{K_{j}}{q_{j}^{*}-1}\bigg(1-\dfrac{K_{i}}{q_{i}^{*}+1}\bigg),
	\end{equation*}
	where $K_{i}$ and $K_{j}$ can be obtained from 	\eqref{eq_kt}. On solving the above equation, we obtain
	\begin{equation*}
		\dfrac{K_{i}}{q_{i}^{*}}-\dfrac{K_{i}}{q_{i}^{*}+1} + \dfrac{K_{j}}{q_{j}^{*}}-\dfrac{K_{j}}{q_{j}^{*}-1}	\leq \dfrac{K_{i}K_{j}}{q_{i}^{*}q_{j}^{*}} - \dfrac{K_{i}K_{j}}{(q_{i}^{*}+1)(q_{j}^{*}-1)}.
	\end{equation*}
	After further modifications, we can rewrite the above equation as 
	\begin{eqnarray*}
		\label{eq:true_inequality_1}
		q_{j}^{*2}K_{i}-q_{j}^{*}(K_{i}+K_{i}K_{j})  \leq  - K_{i}K_{j} +  q_{i}^{*2}K_{j} +  q_{i}^{*} (K_{j}- K_{i}K_{j}).
	\end{eqnarray*}
	In the above equation, we can replace $ (-K_{i}K_{j}+q_{i}^{*2}K_{j} +q_{i}^{*}(K_{j}- K_{i}K_{j}))$ by $C_{1}$, and by rearranging the terms, we get \eqref{LBUB_1}. This completes the proof for the first necessary condition. The second necessary condition for \eqref{express2} can be proved along the similar lines of the above proof to obtain \eqref{LBUB_2}. It can be observed that the two conditions of this theorem are also sufficient because the bounds are obtained by rearranging the terms in the condition on local minima.
	\end{IEEEproof}
\subsection{Low-Complexity List-Decoding Algorithm}	
\label{sec:proposed_method}
For large values of $N$ and $q_{sum}$, it is not feasible to compute the optimal ARQ distribution through exhaustive search. Therefore, by using Theorem \ref{NeccSuff}, we are ready to synthesize a low-complexity algorithm to solve Problem \ref{opt_problem_2}.
\begin{proposition}
\label{prop_1}
If the ARQ distribution $\mathbf{q}$ is chosen such that $\dfrac{q_{j}^{*}}{q_{i}^{*}} = \sqrt{\dfrac{K_{j}}{K_{i}}}$, for $i \neq j$, then $\mathbf{q}$ is a local minima of the search space at a high SNR regime. 
\end{proposition}
\begin{IEEEproof}
	At a high SNR, $\phi$ is very small, and therefore, the product $K_{i}K_{j}$ is negligible. Therefore, we can rewrite \eqref{LBUB_1} and \eqref{LBUB_2} as     
\begin{eqnarray*}
	q_{j}^{*2}K_{i}-q_{j}^{*}K_{i} &\leq &q_{i}^{*2}K_{j} + q_{i}^{*}K_{j}, \\
	q_{j}^{*2}K_{i}+q_{j}^{*}K_{i} &\geq & q_{i}^{*2}K_{j} - q_{i}^{*}K_{j},
\end{eqnarray*}	
respectively. On rearranging the above equations, we get
\begin{eqnarray*}
	q_{j}^{*2}K_{i}-q_{i}^{*2}K_{j} &\leq &q_{j}^{*}K_{i} + q_{i}^{*}K_{j}, \\
	q_{j}^{*2}K_{i}-q_{i}^{*2}K_{j} &\geq & -q_{j}^{*}K_{i} - q_{i}^{*}K_{j},
\end{eqnarray*}	
respectively. Now, by equating $q_{j}^{*}K_{i} + q_{i}^{*}K_{j} = \epsilon_{i,j}$, the above equations become
\begin{eqnarray}
	\label{eq:approx_inequality_1}
	q_{j}^{*2}K_{i}-q_{i}^{*2}K_{j} &\leq & \epsilon_{i,j}, \\
	\label{eq:approx_inequality_2}
	q_{j}^{*2}K_{i}-q_{i}^{*2}K_{j} &\geq & - \epsilon_{i,j},
\end{eqnarray}	
where $\epsilon_{i,j}$ is a positive number dependent on $K_{i}, K_{j}, q_{i}^{*}, q_{j}^{*}$. If we choose $q_{i}^{*}$ and $q_{j}^{*}$ such that $q_{j}^{*2}K_{i}-q_{i}^{*2}K_{j} = 0$, for $i \neq j$, then it ensures that the inequalities in \eqref{eq:approx_inequality_1} and \eqref{eq:approx_inequality_2} are trivially satisfied. Thus, choosing 
\begin{equation}
	\label{eq:succ_bound}
	\dfrac{q_{j}^{*}}{q_{i}^{*}} = \sqrt{\dfrac{K_{j}}{K_{i}}},
\end{equation}
for every pair $i, j \in [N]$ such that $i \neq j$, satisfies the sufficient conditions given in \eqref{eq:approx_inequality_1} and \eqref{eq:approx_inequality_2} at high SNR values. This completes the proof.
\end{IEEEproof}
Based on the results in Proposition \ref{prop_1}, we formulate Problem \ref{relaxed_problem}, as given below, as a means of solving Problem \ref{opt_problem_2} at a high SNR regime. 
	\vspace{0.5cm}
	\begin{mdframed}
		\begin{problem}
			\label{relaxed_problem}
			For a given $\{K_{1}, K_{2}, \ldots, K_{N}\}$ and $q_{sum}$, find $q_{1}^{*},q_{2}^{*},\ldots q_{N}^{*}$ such that $$\frac{q_{j}^{*}}{q_{i}^{*}} = \sqrt{\dfrac{K_{j}}{K_{i}}}, \mbox{where $K_{i}$ is as given in \eqref{eq_kt}}, i,j \in [N] \ \mbox{where} \ i \neq j, $$ $q_{i}^{*} \geq 1, q_{i}^{*}\in \mathbb{Z_{+}}, ~\forall i, \ \sum_{i=1}^{N}q_{i}^{*} = \ q_{sum}.$  
		\end{problem}
	\end{mdframed}
	\vspace{0.5cm}	
	Now, from Problem \ref{relaxed_problem}, it is observed that a solution is not guaranteed because the ratio $\sqrt{K_{j}/K_{i}}$, which is computed based on the LOS components and the SNR, need not be in $\mathbb{Q}$. Therefore, we first propose a method to solve Problem \ref{relaxed_problem} without the integer constraints, i.e., to find an ARQ distribution $\mathbf{q} \in \mathbb{R}^{N}$. This type of problem can be solved by using the system of linear equations of the form $\mathbf{R} \mathbf{q} = \mathbf{s}$ to obtain $\mathbf{q}_{real} = \mathbf{R}^{-1}\mathbf{s},$ where $\mathbf{q} = [q_{1}, q_{2}, \ldots, q_{N}]^{T}$, $\mathbf{s} = [0, 0, \ldots,0, q_{sum}]^{T}$ and $\mathbf{R} \in \mathbb{R}^{N \times N}$ such that $\mathbf{R}(j,j+1) = 1$ for $1 \leq j \leq N$, $\mathbf{R}(N, j) = 1$, for $1 \leq j \leq N$, $\mathbf{R}(j,j) = -r_{j, j+1}$, for $r_{j,j+1} \triangleq \sqrt{K_{j+1}/K_{j}}$ and $1 \leq j \leq N-1$, and the rest of the entries of $\mathbf{R}$ are zeros. Subsequently, using the solution $\mathbf{q}_{real}$ in $\mathbb{R}^{N}$, we propose to obtain a solution in the true search space $\mathbb{S}$ by generating a list as given in Algorithm \ref{List algoritm} \cite{wiopt}. Finally, the ARQ distribution in the list that minimizes $\tilde{p}_{d}^{s}$ would be the final solution. Note that the size of the list captures the number of candidate ARQ distributions shortlisted by our algorithm, and as a result, smaller list size results in fewer number of evaluations using $\tilde{p}_{d}^{s}$, and vice-versa. 

\begin{algorithm}
\caption{\label{List algoritm}List Creation Based Algorithm For CC-HARQ-SF Based Non-Cumulative Network }
\label{LIST_algorithm}
\begin{algorithmic}[1]
		\Require $\mathbf{R}$, $\mathbf{s}$, $q_{sum}$, $\mathbf{c} = [c_{1}, c_{2}, \ldots, c_{N}]$
			\Ensure $\mathcal{L} \subset \mathbb{S}$ - List of ARQ distributions in search space $\mathbb{S}$.
			\State Compute $\mathbf{q}_{real} = \mathbf{R}^{-1}\mathbf{s}$, $\tilde{\mathbf{q}} = \lceil \mathbf{q}_{real} \rceil$.
			\For {$i = 1:N$} 
			\If {$\tilde{q}_{i} = 0$} 
			\State $\tilde{q}_{i} = \tilde{q}_{i} + 1$			
			\EndIf
			\EndFor
			\State Compute $E = \left(\sum_{i = 1}^{N} \tilde{q}_{i}\right) - q_{sum}$
			\State $\mathcal{L} = \{\mathbf{q} \in \mathbb{S} ~|~ d(\mathbf{q}, \tilde{\mathbf{q}}) = E, q_{j} \ngtr q_{i} \text{ for }  c_{i} < c_{j}\}$. 		
		\end{algorithmic}
	\end{algorithm}
	\begin{figure}[h!]
		\centering \includegraphics[scale = 0.5]{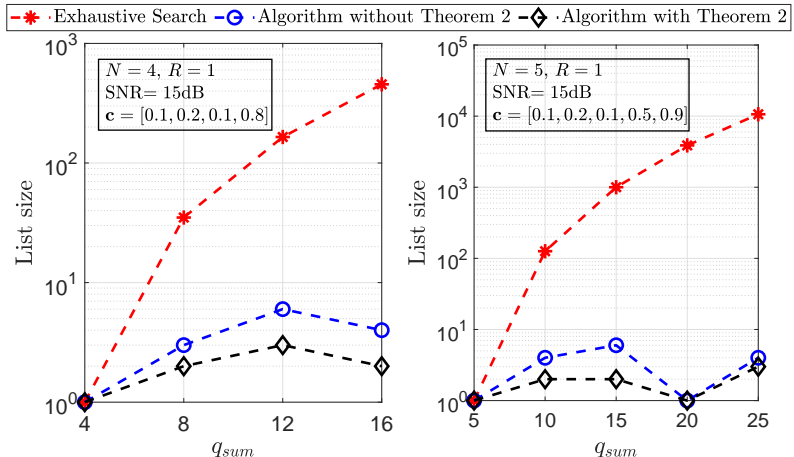}
		\vspace{-0.5cm}	
		\centering{\caption{Comparison of list size between exhaustive search and the proposed low-complexity algorithm with and without Theorem \ref{LOShighlow_SF} for CC-HARQ-SF based non-cumulative network.
				\label{fig:List_size}}}
	\end{figure}
\begin{figure*}[h]
\begin{center}
\includegraphics[scale = 0.5]{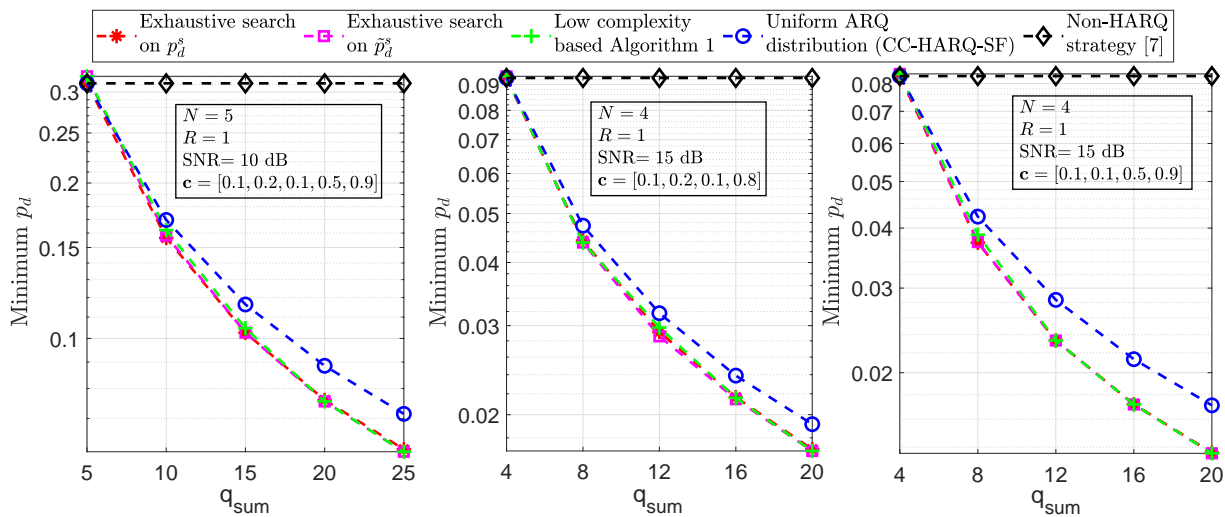}
\vspace{-0.3cm}
{\caption{PDP comparison in CC-HARQ-SF based non-cumulative network between (i) exhaustive search using original PDP expression, $p_{d}^{s}$, (ii) exhaustive search using approximated PDP expression, $\tilde{p}_{d}^{s}$, (iii) proposed low-complexity algorithm (given in Algorithm \ref{LIST_algorithm}), and (iv) uniform ARQ distribution along with ARQ distribution in a non-HARQ strategy (Type-1 ARQs) \cite{our_work_TWC_1}. 
					\label{fig:pdp_comparison}}}
		\end{center}
	\end{figure*}	

	\subsection{Simulation Results for a CC-HARQ-SF based Non-Cumulative Network}
	\label{sec:Sims}
	In this section, we present simulation results to validate our theoretical analysis and to showcase the benefits of using the CC-HARQ-SF protocol over a Type-1 ARQ based multi-hop model. On the one hand, the computational complexity for solving Problem \ref{opt_problem_SF} and Problem \ref{opt_problem_2} using an exhaustive search is ${q_{sum}-1}\choose{N-1}$. On the other hand, the computational complexity of our method is determined by the complexity of computing the inverse of a matrix and that of Algorithm \ref{LIST_algorithm}. To compare the complexity of the two methods, we plot their list size in Fig. \ref{fig:List_size} for several values of $q_{sum}$ with $N = 4$ and $N = 5$. We plot the list size both with and without incorporating the results of Theorem \ref{LOShighlow_SF}. When using Theorem \ref{LOShighlow_SF}, by subtracting one ARQ from all possible $\binom{N}{E}$ positions from $\tilde{\mathbf{q}}$, we discard those ARQ distributions which follow the rule $q_{i}>q_{j}$ whenever $c_{i} > c_{j}$. As a result, we see that the list size shortens after incorporating the rule of Theorem \ref{LOShighlow_SF} for $N = 5$. Based on the simulation results, we observe that the ARQ distribution, which minimizes the PDP from the list $\mathcal{L}$ matches the result of exhaustive search.
	
	Finally, to showcase the benefits of using CC-HARQ-SF in a multi-hop network, we plot the minimum PDP against several values of $q_{sum}$ in Fig. \ref{fig:pdp_comparison} by using (i) an exhaustive search on $p^{s}_{d}$ (given in \eqref{eq:pdp_expression_SF}), (ii) an exhaustive search on $\tilde{p}^{s}_{d}$ (given in \eqref{eq:Approx_pdp_expression}), (iii) proposed low-complexity algorithm, (iv) uniform ARQ distribution in CC-HARQ-SF based strategy and (v) non-HARQ strategy \cite{our_work_TWC_1}. There are mainly two observations from the simulation results in Fig. \ref{fig:pdp_comparison}: (a) the PDP improves in the case of optimal ARQ distribution (with an exhaustive search on both $p_{d}^{s}$ and $\tilde{p}_{d}^{s}$) over uniform ARQ distribution and non-hybrid based ARQ distribution, and (b) our proposed approximation $\tilde{p}_{d}^{s}$ produces reasonably accurate results.
	\section{CC-HARQ-SF Based Fully-Cumulative Network}
	\label{sec:fully_cumm_SF}
	In the CC-HARQ-SF based non-cumulative network, the nodes in the network do not have the knowledge on the ARQs given to the other nodes, thereby leading to wastage of ARQ resources. Observing this limitation of the CC-HARQ-SF based non-cumulative network, in this section, we consider a setup where every node is aware of the ARQs assigned to the node before it, unlike the non-cumulative model. Therefore, the following node can use the unused ARQs from the preceding node. In addition, we also assume that the packet structure contains a particular portion, referred to as the \emph{counter}, in order to carry the number of ARQs not utilized by the preceding nodes in the network. We emphasize that the need to include the counter in the packet structure has a vital role since a given relay node can only hear the residual ARQs from the node immediately preceding one \cite[Section V]{our_work_TWC_1}. When using a counter, every node can use the residual ARQs from all its preceding nodes in addition to its ARQs, thus reducing the end-to-end PDP. Henceforth, throughout the paper, we refer to this multi-hop network as the CC-HARQ-SF based fully-cumulative network.   
	
\begin{figure*}[h!]
		\centering \includegraphics[scale = 0.50]{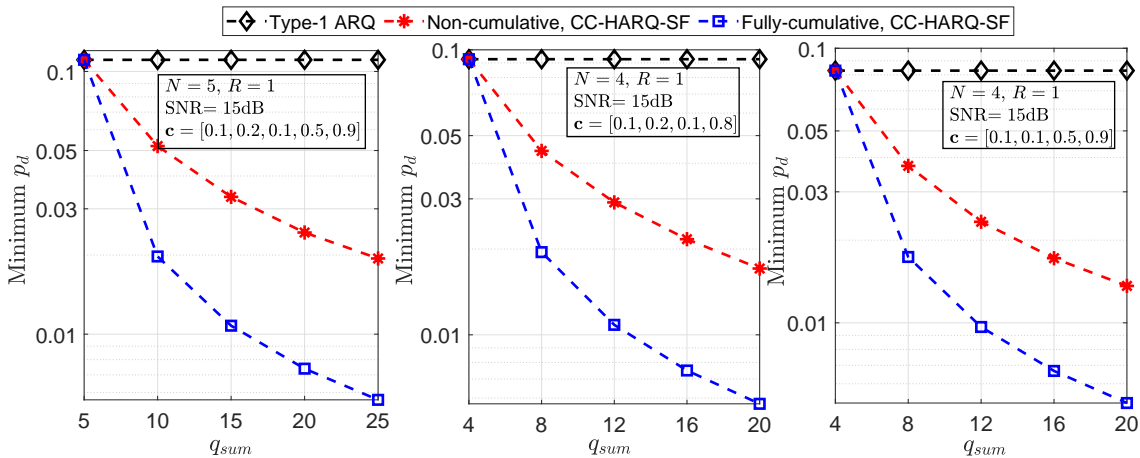}
		\vspace{-0.4cm}
		\centering{\caption{Comparison of the minimum PDP, denoted by $p_{d}$, for : (i) CC-HARQ-SF based non-cumulative network, (ii)  CC-HARQ-SF based fully-cumulative network, and (iii) Type-1 ARQ strategy.
				\label{fig:fully_cumm_SF}}}
	\end{figure*}

	To explain the model further, assume that on the first link, $q^{'}_{1}$ out of $q_{1}$ ARQs are used to successfully transmit the packet to the next node by implementing CC-HARQ-SF, and as a consequence, $q_{1}^{''}= q_{1}-q_{1}^{'}$ ARQs are not utilised. Since the following node in the chain is aware of both $q_{1}$ ARQs and the number of attempts made by the preceding node for the successful transmission of the packet, it can use $q_{2} +q_{1}^{''}$ ARQs to transmit the packet to its successor node. Furthermore, to assist the successor node in the chain to make aware of any residual number of ARQs, the second node encodes the number $q_{2}+q_{1}^{''}$ in the counter and then starts transmitting the packet to the successor node. If the second node uses $q_{2}^{'}$ attempts, then the third node can use $q_{3} + q_{2}+q_{1}^{''}-q_{2}^{'}$ number of attempts. Similarly, the third node encodes the number $ q_{3}+q_{2}+q_{1}^{''}-q_{2}^{'}$ in the counter and then starts transmitting the packet to its successor node. This way, each intermediate relay node potentially gains additional ARQs for packet transmission. We emphasize that every relay node updates the counter in the packet only once, and as a result, the additional delay incurred by this update process is negligible. In this scheme, note that as the preceding node may have more ARQs than allotted to it (because of additional ARQ accumulated from its preceding node), the next node in the chain does not know the maximum number of attempts that would be made by its preceding node. Therefore, a given node, upon sending a NACK to its preceding node, will wait for the packet only for a specified amount of time, say $\tau$ seconds. If it does not receive the packet within $\tau$ units of time, then it implies that the preceding node has exhausted the allotted number of ARQs, and therefore, the packet is said to be dropped in the network. It is intuitive that this process of cumulatively adding the unused ARQs at each hop will reduce the PDP in comparison with the non-cumulative strategy without changing the sum constraint $q_{\text{sum}} = \sum_{i=1}^{N} q_{i}$. To showcase the benefits, we have shown the simulation results for $4$-hop and $5$-hop networks in Fig. \ref{fig:fully_cumm_SF} where it can be observed that the CC-HARQ-SF based fully-cumulative network outperforms the CC-HARQ-SF based non-cumulative network. 
	
Now, we present the result on the optimal ARQ distribution for the CC-HARQ-SF based fully-cumulative network.
	\begin{theorem}
		\label{thm:Fully_cumm_slowfading}
		For a given ARQ distribution $\mathbf{q} = [q_{1}, q_{2}, \ldots, q_{N}]$ and $q_{sum}$ in a CC-HARQ-SF based fully-cumulative network, the optimal ARQ distribution can be given by $[q_{sum}, 0, \ldots, 0]$.
	\end{theorem}
	\begin{IEEEproof}
		First, we want to prove the result for $N=2$. With ARQ distribution $[q_{1},q_{2}]$, the PDP is
		\begin{equation}
			pdp_{2}^{s} = P_{1q_{1}}^{s} + \sum_{i=0}^{q_{1}-1} (P_{1i}^{s} -P_{1(i+1)}^{s} )P_{2(q_{2}+q_{1}-(i+1))}^{s}, 	
		\end{equation}
		where $P_{10}^{s}=1$. Let us transfer one ARQ from the second link to the first link by considering that $q_{2}>1$. Therefore, the updated PDP expression in this case is $pdp_{2}^{s'} = P_{1(q_{1}+1)}^{s}  + \sum_{i=0}^{q_{1}} (P_{1i}^{s} -P_{1(i+1)}^{s} )P_{2(q_{2}+q_{1}-(i+1))}^{s}.$ On calculating $pdp_{2}^{s} - pdp_{2}^{s'}$, we get $pdp_{2}^{s} - pdp_{2}^{s'} = (P_{1q_{1}} - P_{1(q_{1}+1)}^{s} )(1-P_{2(q_{2}-1)}^{s})$. It can be observed that $P_{1q_{1}}^{s}  > P_{1(q_{1}+1)}^{s} $ and $(1-P_{2(q_{2}-1)}^{s} ) \geq 0$, because $P_{2(q_{2}-1)}^{s}  < 1$. Therefore, we conclude that the above equation results in a positive number and hence, $pdp_{2}^{s} > pdp_{2}^{s'}$. Furthermore, if we keep transferring one ARQ from the second link to the first link and calculating the difference between the old PDP and the new PDP (after transferring one ARQ), the difference is always positive. In this way, the optimal ARQ distribution can be given by $[q_{1}+q_{2}, 0]$. This completes the proof for $N=2$. Now, by using the hypothesis step of induction, let us assume that the result is true for $N=k$. To prove the result for $N=(k+1)$, let $\mathbf{q} = [q_{1},q_{2}, \ldots, q_{k+1}]$ be the ARQ distribution across the $k+1$ nodes. Since, the nodes are applying the  fully-cumulative scheme, we assume that the set of nodes from the $2^{nd}$ to the $(k+1)$-th node as a single virtual node. Henceforth, we refer to this virtual node as $R_{v}$. After forming a virtual node, now, we have a network comprising a source node, $R_{v}$ and a destination node. Also, the $R_{v}$ node has been given a total of $q_{sum,v} = \sum_{i=2}^{k+1}q_{i}$ number of ARQs and $q_{sum,v}$ is internally distributed among the $k$ nodes constituting the virtual node, $R_{v}$. Let $PDP_{v}^{s}(q_{sum,v})$ represent the PDP at the virtual node. Now, the PDP expression for this above mentioned two-hop network can be written as 
		\begin{equation}
			\label{virtual_pdp_2h_sf}	
			PDP_{2v}^{s} = P_{1q_{1}}^{s}  + \sum_{i=0}^{q_{1}-1}(P_{1i}^{s} -P_{1(i+1)}^{s} )PDP_{v}^{s}(q_{sum,v}+q_{1}-(i+1)), 
		\end{equation} 
		where $P_{10}^{s}=1$. Now, in the above equation, it can be observed that for the fixed $q_{1}$, $P_{1i}^{s}$ and $P_{1(i+1)}^{s}$ for each iteration $`i'$, the PDP expression can be minimized by minimizing the $PDP_{v}^{s}(q_{sum,v}+q_{1}-(i+1))$. Also, by using the assumption on $N=k$ from the induction step, we know that $[q_{sum}, 0 , \ldots, 0]$ will minimize $PDP_{v}^{s}$. Furthermore, for \eqref{virtual_pdp_2h_sf}, we have the ARQ distribution $[q_{1}, q_{sum,v}, 0 , \ldots, 0]$. In order to find the optimal ARQ distribution of \eqref{virtual_pdp_2h_sf}, we use similar approach as for $N=2$. That is, we start transferring one ARQ from the virtual node to the first link and write the new PDP expression as 
		\begin{eqnarray*}
			PDP_{2v}^{s'} = P_{1(q_{1}+1)}^{s}  + \sum_{i=0}^{q_{1}} (P_{1i}^{s}  - P_{1(i+1)}^{s} )PDP_{v}^{s}(q_{sum,v}+q_{1}-(i+1)).
		\end{eqnarray*}    
		On calculating $PDP_{2v}^{s}-PDP_{2v}^{s'}$, we get $PDP_{2v}^{s}-PDP_{2v}^{s'} =  (P_{1q_{1}}^{s} -P_{1(q_{1}+1)}^{s} ) (1-PDP_{v}^{s}(q_{sum,v}-1))$. It can be observed that $0 \leq P_{1q_{1}}^{s} -P_{1(q_{1}+1)}^{s}  \leq 1$ and $(1-PDP_{v}^{s}(q_{sum,v}-1)) \geq 0$, therefore, it implies $PDP_{2v}^{s} \geq PDP_{2v}^{s'}$. Hence, on transferring one ARQ from the virtual node to the first link, the PDP will decrease. Therefore, we can conclude that the optimal ARQ distribution is $[q_{sum}, 0, \ldots, 0]$.  
	\end{IEEEproof} 	
	\section{Delay Analysis for CC-HARQ Strategies in Slow-Fading Scenarios}
	\label{sec:Sims_delay_analysis_SF}
\begin{figure}[h!]
		\centering \includegraphics[scale=0.47]{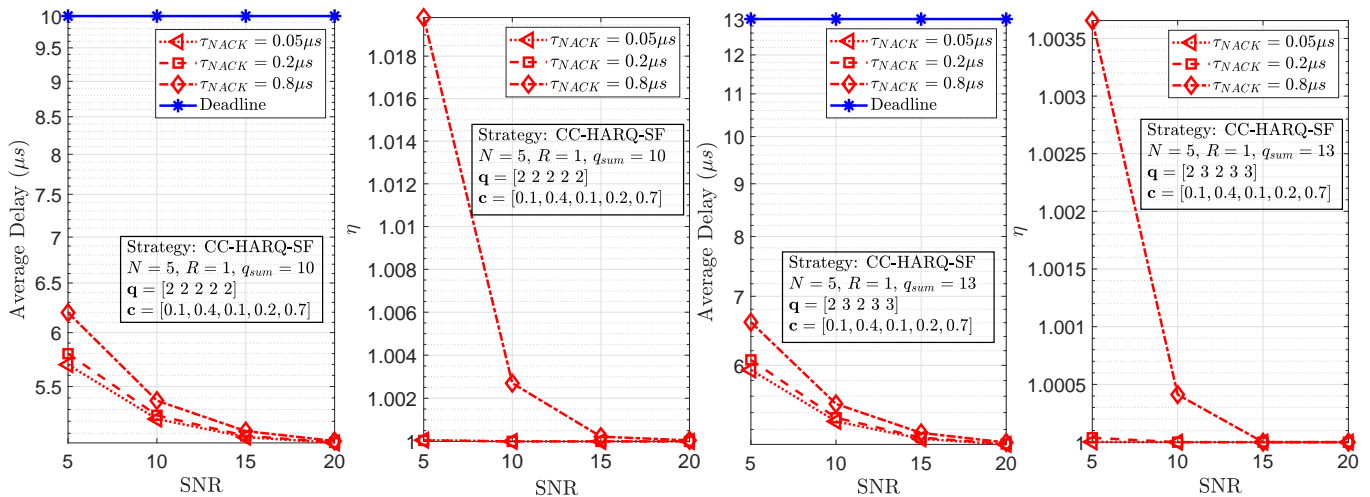}
		\vspace{-0.8cm}	
		\centering{\caption{Average delay on the packets and the deadline violation parameter ($\eta$) for various $\tau_{NACK}$ in CC-HARQ-SF based strategies.
				\label{fig:average_delay_SF}}}
	\end{figure}

	In this section, we present a detailed analysis on end-to-end delay for CC-HARQ-SF strategies. The objective of this delay analysis is to show that the packets in a non-cumulative network have a high probability of reaching the destination before the deadline, whenever the delay overheads from ACK/NACK are sufficiently small. To demonstrate the results, we obtain $q_{sum}$ as $\floor{\frac{\tau_{total}}{\tau_{p}+\tau_{d}}}$ without considering the resources for ACK/NACK in the reverse channel, where $\tau_{total}, \tau_{d}$ and $\tau_{p}$ are as defined in Section \ref{sec:system_model}. Subsequently, we introduce a different resolution of delays from NACK, say $\tau_{NACK}$ time units, and then observe the impact on the end-to-end delay on the packets. Assuming $\tau_{p} + \tau_{d} = 1 ~\mu s$, we set the deadline for end-to-end packet delay as $q_{sum} ~\mu s$. Then, by sending an ensemble of $10^6 $ packets to the destination through the CC-HARQ-SF strategy, we compute the following metrics when $\tau_{NACK} \in \{0.05, 0.2, 0.8\}$ in $\mu s$: (i) the number of packets that were dropped in the network (denoted by $P_{Drop}$) due to insufficient ARQs at the intermediate nodes, (ii) the number of packets that reach the destination after the deadline (denoted by $P_{Deadline}$), and finally, (iii) the average end-to-end delay on the packets, which is computed by tracking the end-to-end delay incurred on each packet and then averaging them empirically. The average delay metric is plotted in Fig. \ref{fig:average_delay_SF} for various values of $\mbox{SNR}$ for a specific value of $N$ and the LOS vector $\mathbf{c}$.
	The plots suggest that the average delay is significantly lower than that of the deadline, especially when $\tau_{NACK}$ is small, owing to the opportunistic nature of CC-HARQ-SF strategies. However, as $\tau_{NACK}$ increases, the average delay is pushed slightly closer to the deadline. Furthermore, to capture the behaviour of deadline violations due to higher $\tau_{NACK}$, in Fig. \ref{fig:average_delay_SF}, we also plot $\eta = \frac{P_{Drop} + P_{Deadline}}{P_{Drop}}$, which is a metric used to represent the ratio of packets that reach
the destination after the given deadline as compared to the dropped packets. The plots confirm that when $\tau_{NACK}$ is sufficiently small compared to $\tau_{p} + \tau_{d}$ (see $\tau_{NACK} = 0.05 ~\mu s$ at \mbox{SNR} = 10, 15, 20 dB), the packets that reach the destination arrive within the deadline with an overwhelming probability as $\eta = 1$.
	
\begin{figure}[h!]
		\centering \includegraphics[scale=0.5]{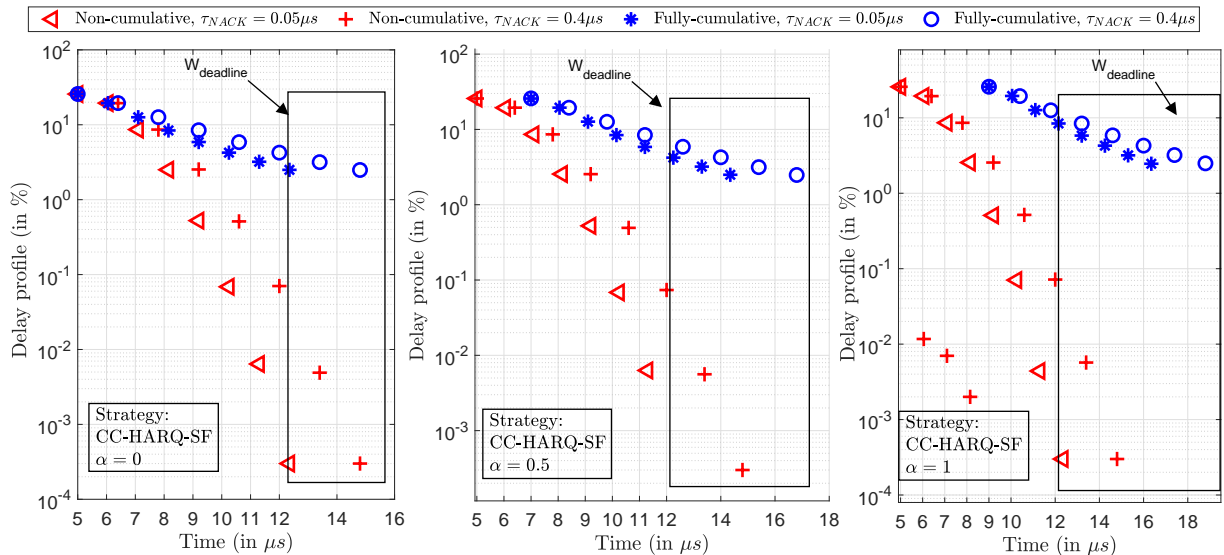}
		\vspace{-0.4cm}	
		\centering{\caption{Simulation results on delay profiles for CC-HARQ-SF based strategies using a $5$-hop network with $\mathbf{c}= [0.1, 0.5, 0.1, 0.3, 0.7]$ and $q_{sum}= 12$ at rate $R=1$ and SNR$=5$dB with $10^{6}$ packets wherein some percentage of packets are dropped either due to outage and some percentage are dropped due to deadline violation, denoted by $W_{\text{deadline}}$ (marked in the rectangle). 
		\label{fig:Delay_profile_SF}}}
	\end{figure}

If we accurately incorporate every delay parameter introduced by the relay nodes into $q_{sum}$, then all the packets that reach the destination arrive within a given deadline with probability one. Therefore, only the PDP metric is sufficient to study the proposed strategies. However, if we do not incorporate some delays, or if we incorporate lower values of delays in $q_{sum}$ due to inaccuracy in measurements, then some packets may still reach the destination beyond the deadline. Therefore, in such cases, non-incorporation of delays in $q_{sum}$ must be studied in the form of deadline violation in addition to the PDP metric. In order to study one such practical use-case of not incorporating delay, we discuss the implementation of crypto-primitives at each relay node in order to provide the confidentiality feature to the counter in the packets.
As a result, when the packet is successfully decoded at the next node, it needs to decrypt it by using an appropriate crypto-primitive. Since this procedure results in an additional processing delay on the packet, we represent this delay by $T_{c}~\mu s$. We assume that the delay introduced on the packet per hop for each transmission is $T = \tau_{p}+\tau_{d}= 1 ~\mu s$, and we exclude the time for executing crypto-primitives when calculating $q_{sum}$. Now, we analyse the effect of crypto-primitives on end-to-end delay by choosing $T_{c} = \alpha T$, where $\alpha = 0, 0.5, \mbox{ and }1$. Note that different values of $\alpha$ are used to capture the throughputs (or the speed) of different architectures for the crypto-primitives. In Fig. \ref{fig:Delay_profile_SF}, we have shown the effect of $\alpha$ on the end-to-end delay by plotting the delay profiles (in percentage) of both non-cumulative and fully-cumulative strategies. In this context, delay profile is defined as the distribution providing the percentage of packets that reach the destination at a particular end-to-end delay. To be precise, when plotting the delay profile, x-axis represents the set of all end-to-end delay numbers with which the packets reach the destination, and the y-axis represents the percentage of packets that reached the destination at that end-to-end delay. For generating the plots, we have used an ensemble of $10^{6}$ packets and we considered $N=5$, $R=1$, SNR$=5$ dB, $q_{sum}=12$, and $\mathbf{c} = \{0.1, 0.5, 0.1, 0.3, 0.7\}$. It can be noted that the percentage of packets that reach the destination after the given deadline can be treated under the category of deadline violation. Evidently, delay profiles are the same for the non-cumulative strategy irrespective of the value of $\alpha$ because there is no counter present in it. However, it can be observed in a fully-cumulative network that as $\alpha$ increases, the percentage of packets violating the deadline (where the deadline is $12 ~\mu s$) increases. It can also be visualized that as $\alpha$ increases, the width of the rectangle increases (wherein the rectangle shows the percentage of packets violating the deadline).
	\begin{figure}[h!]
		\centering \includegraphics[width=18cm, height=10.5cm]{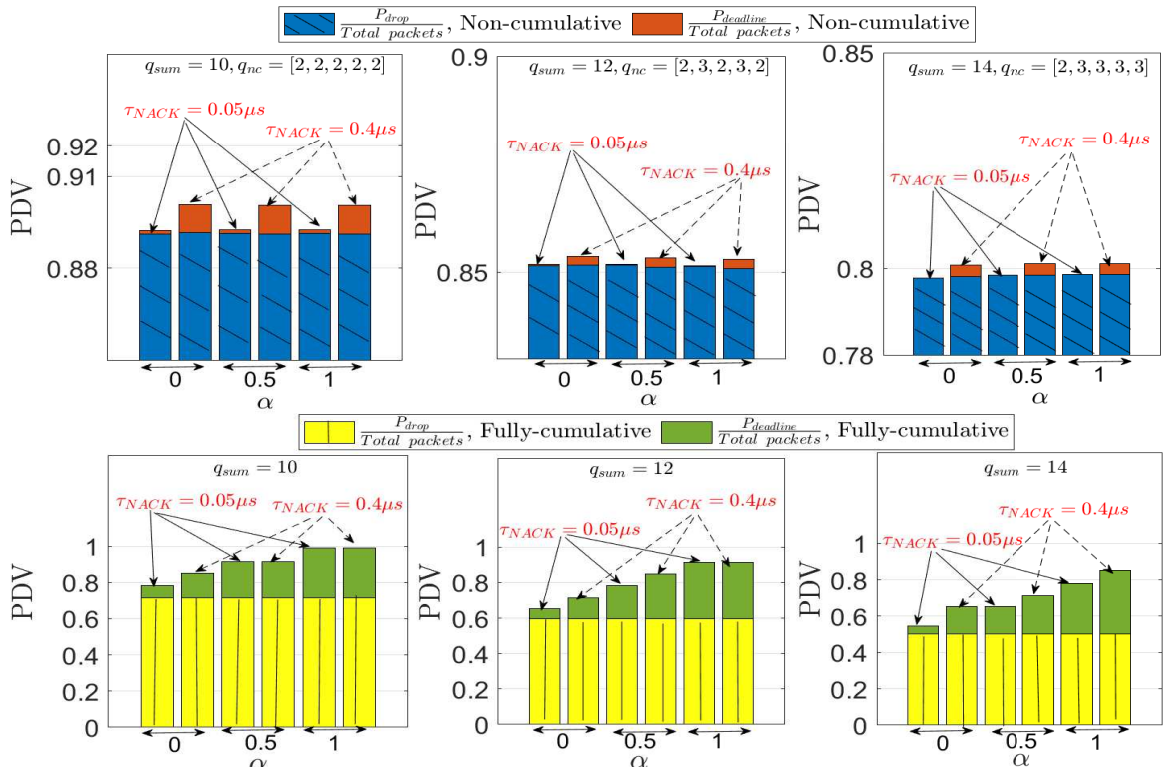}
		\vspace{-0.5cm}	
		\centering{\caption{Illustration on PDV for the non-cumulative and fully-cumulative networks while implementing CC-HARQ-SF strategies for different $\tau_{NACK}$ and $\alpha$ values.  
				\label{fig:PDV_SF}}}
	\end{figure}Furthermore, if the effect of $\alpha$ is not considered when designing $q_{sum}$, when $\alpha = 0$, then there is a non-zero probability that some packets may reach the destination beyond the deadline; however it is very small (see the first plot in Fig.  \ref{fig:Delay_profile_SF}).
	
	Now, in addition to PDP, we define a new metric referred to as probability of deadline violation (PDV), which can be defined as $\text{PDV} = \frac{P_{Drop}+P_{Deadline}}{\text{Total packets}}$. In Fig. \ref{fig:PDV_SF}, we plot the PDV 
	for our CC-HARQ-SF based non-cumulative and fully-cumulative networks as a function of $\alpha$ for a $5$-hop network with different $q_{sum}$. In the plots, $\mathbf{q_{nc}}$ represents the ARQ distribution taken for generating the plots for the non-cumulative strategy and $\{q_{sum},0,\dots,0\}$ is the ARQ distribution for the fully-cumulative network. Also, for a given $q_{sum}$, the deadline for packets to reach the destination is $q_{sum} ~\mu s$  by assuming that $T = 1 ~\mu s$. The plots confirm that: (i) PDV of the non-cumulative network do not change with $\alpha$, (ii) PDV of the fully-cumulative network increases with increasing values of $\alpha$; this is because $N-1$ nodes made use of the counter in the packet, thereby adding a significant delay of $(N-1)T_{c} ~\mu s$  to the packet.
	
	\section{Optimal ARQ Distribution in CC-HARQ-FF based Non-Cumulative Network}
	\label{sec:analysis_FF}
	In this section, we analyse the optimal ARQ distribution for the CC-HARQ-FF network wherein the channels across the attempts are assumed to be statistically independent, as given in Fig. \ref{fig:system_model}(b). Along the similar lines of Section \ref{sec:analysis_SF}, for a CC-HARQ-FF based non-cumulative network, the PDP expression is
	\begin{equation}
		\label{eq:pdp_expression_FF}
		p_{d}^{f} = P_{1q_{1}}^{f} + \sum_{k=2}^{N} P_{kq_{k}}^{f}\Bigg( \prod_{j=1}^{k-1}  (1-P_{jq_{j}}^{f})\Bigg),
	\end{equation}
	where $P_{kq_{k}}^{f} = \mbox{Pr} \left(R > \log_{2}(1+ \sum_{j=1}^{q_{k}}|h_{kj}|^{2} \gamma)\right)$. Note that when $q_{k}=0$, we have $\sum_{j=1}^{q_{k}}|h_{kj}|^{2} \gamma =0$ which implies $P_{k0}^{f}=\mbox{Pr} \left(R > 0\right) = 1$, for $k \in [N]$. Now, we are interested in allocating the ARQs to the nodes to minimize the PDP in \eqref{eq:pdp_expression_FF} for a given sum constraint $\sum_{i = 1}^{N} q_{i} = q_{sum}$. Therefore, we formulate our optimization problem in Problem \ref{opt_problem_FF}, as shown below. Throughout this paper, we refer to the solution of Problem \ref{opt_problem_FF} as the optimal ARQ distribution for the CC-HARQ-FF based non-cumulative network.        

	\begin{mdframed}
		\begin{problem}
			\label{opt_problem_FF}
			For a given $\mathbf{c}$, $\gamma = \frac{1}{\sigma^{2}}$ and $q_{sum}$, solve $q_{1}^{*},q_{2}^{*},\ldots q_{N}^{*}=\arg\underset{q_{1},q_{2},\ldots q_{N}}{\text{min}}\ p_{d}^{f}$, \text{subject to}, $q_{k} \geq 1, q_{k}\in \mathbb{Z_{+}} \ \forall k \in [N],
				\sum_{i=1}^{N}q_{i} = \ q_{sum}$.  		
		\end{problem}
	\end{mdframed}
	
	It can be observed that $P_{kq_{k}}^{f}$ is dependent on $q_{k}$-th order Marcum-Q function and tackling a higher-order Marcum-Q function analytically is challenging because it contains a lower incomplete gamma function. Therefore, we first present an approximation on the higher-order Marcum-Q function at high SNR regimes.	
	\subsection{Approximation on the $q_{k}$-th Order Marcum-Q Function}
	\begin{theorem}
		\label{theorem:approx_Marcum_Q_FF}
		At high SNR regime, we can write the generalized Marcum-Q function of order $q_{k}$, denoted by $Q_{q_{k}}(\mathbb{a}_{k},b_{k})$, as $Q_{q_{k}}(\mathbb{a}_{k},b_{k})= \tilde{Q}_{q_{k}}(\mathbb{a}_{k},b_{k}) + O((b_{k}^{2q_{k} + 2}))$, where $\tilde{Q}_{q_{k}}(\mathbb{a}_{k},b_{k}) \triangleq 1 - \frac{1}{q_{k}!}\big(\frac{b_{k}^{2}}{2}\big)^{q_{k}}e^{\frac{-\mathbb{a}_{k}^{2}}{2}}$ for all $k \in [N]$, such that $\mathbb{a}_{k} =\sqrt{\frac{2q_{k}c_{k}}{(1-c_{k})}}$ and $b_{k}= \sqrt{\frac{2(2^{R}-1)}{\gamma(1-c_{k})}}$. Furthermore, at high SNR, we have $Q_{q_{k}}(\mathbb{a}_{k},b_{k}) \approx \tilde{Q}_{q_{k}}(\mathbb{a}_{k},b_{k})$. 
	\end{theorem}
	\begin{IEEEproof} 
		In the context of the $N$-hop network discussed in the previous section, the generalized Marcum-Q function of order $q_{k}$ associated with the $k$-th hop is given by 
		\begin{eqnarray}
			\label{generalizedMarcumExpression}
			Q_{q_{k}}(\mathbb{a}_{k},b_{k}) = 1 - e^{-\frac{\mathbb{a}_{k}^{2}}{2}}\sum_{i=0}^{\infty}\frac{1}{i!}\bigg(\frac{\Lambda(q_{k}+i , \frac{b_{k}^2}{2})}{\Gamma(q_{k}+i)}\bigg) \bigg(\frac{\mathbb{a}_{k}^{2}}{2}\bigg)^{i},  
		\end{eqnarray}
		where $\Lambda(\cdot,\cdot)$ is a lower incomplete gamma function, $\Gamma(\cdot)$ is a gamma function defined as $\Gamma(n) =(n-1)!, $ $\mathbb{a}_{k}= \sqrt{\frac{2q_{k}c_{k}}{(1-c_{k})}}$, and $b_{k}= \sqrt{\frac{2(2^{R}-1)}{\gamma(1-c_{k})}}$ such that $\gamma = \frac{1}{\sigma^{2}}$. 
		Furthermore, to simplify $\Lambda(\cdot,\cdot)$, we use the Holomorphic extension of incomplete gamma function given by 
		\begin{eqnarray}
			\label{incompleteGamma}
			\Lambda\bigg(q_{k}+i,\frac{b_{k}^2}{2}\bigg) = \Gamma(q_{k}+i)e^{-\frac{b_{k}^{2}}{2}}\bigg(\frac{b_{k}^{2}}{2}\bigg)^{q_{k}}\sum_{j=0}^{\infty}\frac{\big(\frac{b_{k}^{2}}{2}\big)^{i+j}}{\Gamma(q_{k}+i+j+1)}. 
		\end{eqnarray}
		By using \eqref{generalizedMarcumExpression} and \eqref{incompleteGamma}, we can write 
		\begin{eqnarray*}
			Q_{q_{k}}(\mathbb{a}_{k},b_{k}) = 1 - e^{-\frac{\mathbb{a}_{k}^{2}}{2}}e^{-\frac{b_{k}^{2}}{2}}\bigg(\frac{b_{k}^{2}}{2}\bigg)^{q_{k}}\sum_{i=0}^{\infty}\frac{1}{i!}\ \bigg(\frac{\mathbb{a}_{k}^{2}}{2}\bigg)^{i}\sum_{j=0}^{\infty}\frac{ \big(\frac{b_{k}^{2}}{2}\big)^{i+j}}{\Gamma(q_{k}+i+j+1)}. 
		\end{eqnarray*}
		On expanding $e^{-\frac{b_{k}^{2}}{2}}$ using the Taylor series, similar to Theorem \ref{thm:approximation_Marcum_Q_1}, we can write $Q_{q_{k}}(\mathbb{a}_{k},b_{k}) = \tilde{Q}_{q_{k}}(\mathbb{a}_{k},b_{k}) + \epsilon_{Q_{q_{k}}}$, where
\begin{eqnarray}
\label{eq:approximation_marcum_ff}
\tilde{Q}_{q_{k}}(\mathbb{a}_{k},b_{k}) & \triangleq & 1 - \frac{1}{q_{k}!}\bigg(\frac{b_{k}^{2}}{2}\bigg)^{q_{k}}e^{-\frac{\mathbb{a}_{k}^{2}}{2}}, \\
\epsilon_{Q_{q_{k}}} &=&  e^{-a_{k}^{2}/2} \Bigg(\frac{b_{k}^{2}}{2}\Bigg)^{q_{k}}\Bigg[\Bigg(\sum_{i=0}^{\infty} \frac{(-b_{k}^{2}/2)^{i}}{i!}\Bigg)\Bigg(\sum_{i=0}^{\infty} \frac{(a_{k}^{2}/2)^{i}}{i!} \sum_{j=1}^{\infty} \frac{(b_{k}^{2}/2)^{i+j}}{\Gamma (q_{k}+i+j+1)} \Bigg) + \nonumber \\ & &  \frac{1}{q_{k}!}\Bigg(\sum_{i=1}^{\infty}\frac{(-b_{k}^{2}/2)^{i}}{i!}\Bigg) \Bigg]. \nonumber
\end{eqnarray}

Since $\epsilon_{Q_{q_{k}}}$ contains terms of the form $(b_{k}^{2}/2)^{l}$, where $l \in \{q_{k}+1, q_{k}+2, \ldots, \infty\}$, we can write $Q_{q_{k}}(\mathbb{a}_{k},b_{k}) = \tilde{Q}_{q_{k}}(\mathbb{a}_{k},b_{k}) + O(b^{2q_{k} + 2}_{k})$ by considering the dominant term of $\epsilon_{Q_{q_{k}}}$. Note that since $\tilde{Q}_{q_{k}}(\mathbb{a}_{k},b_{k})$ contains $b^{2q_{k}}_{k}$, the term $O(b^{2q_{k} + 2}_{k})$ provides negligible contribution to $Q_{q_{k}}(\mathbb{a}_{k},b_{k})$ when $b_{k}$ is small. Therefore, we have $Q_{q_{k}}(\mathbb{a}_{k},b_{k}) \approx \tilde{Q}_{q_{k}}(\mathbb{a}_{k},b_{k})$ at high SNR values. This completes the proof.
	\end{IEEEproof}
Now, by using the approximation in \eqref{eq:approximation_marcum_ff}, we can approximate $P_{kq_{k}}^{f}$ in \eqref{eq:pdp_expression_FF} as
	\begin{equation}
		\label{eq:approximation_P_k_ff}
		\tilde{P}_{kq_{k}}^{f}  = 1- \tilde{Q}_{q_{k}}(\mathbb{a}_{k},b_{k}) \triangleq\frac{1}{q_{k}!}\bigg(\frac{b_{k}^{2}}{2}\bigg)^{q_{k}}e^{-\frac{\mathbb{a}_{k}^{2}}{2}} = \frac{1}{q_{k}!} \bigg(\frac{\phi}{1-c_{k}}\bigg)^{q_{k}} \bigg(e^{\frac{-a_{k}^{2}}{2}}\bigg)^{q_{k}},
	\end{equation}
	where $\phi = \frac{2^{R}-1}{\gamma}$, and $\tilde{P}_{kq_{k}}^{f}$ denotes the approximation on $P_{kq_{k}}^{f}$. Also, $\mathbb{a}_{k}^{2}=q_{k}a_{k}^{2} $ (where $a_{k}$ is defined earlier in Theorem \ref{thm:approximation_Marcum_Q_1}). Finally, by using \eqref{eq:approximation_P_k_ff}, we can write the approximate version of \eqref{eq:pdp_expression_FF} as
	\begin{equation}
		\label{eq:Approx_pdp_expression_ff}
		\tilde{p}_{d}^{f} = \tilde{P}^{f}_{1q_{1}} + \sum_{k=2}^{N} \tilde{P}^{f}_{kq_{k}} \Bigg( \prod_{j=1}^{k-1}  (1-\tilde{P}^{f}_{jq_{j}})\Bigg),
	\end{equation} 
	where $\tilde{p}^{f}_{d}$ denotes the approximate version of $p^{f}_{d}$. Henceforth, using the approximated expression on $p^{f}_{d}$, we formulate an optimization problem in Problem \ref{approx_opt_problem_2_ff}. Throughout this paper, we refer to the solution of Problem \ref{approx_opt_problem_2_ff} as the near-optimal ARQ distribution for CC-HARQ-FF based non-cumulative network since the objective function in Problem \ref{approx_opt_problem_2_ff} is an approximate version of the objective function in Problem \ref{opt_problem_FF}. 
	
	\begin{mdframed}
		\begin{problem}
			\label{approx_opt_problem_2_ff}
			For a given $\mathbf{c}$, a high SNR $\gamma = \frac{1}{\sigma^{2}}$ and $q_{sum}$, solve $q_{1}^{*},q_{2}^{*},\ldots q_{N}^{*}=\arg\underset{q_{1},q_{2},\ldots q_{N}}{\text{min}}\ \tilde{p}^{f}_{d}$, \text{subject to} $\ q_{k} \geq 1, q_{k}\in \mathbb{Z_{+}} \ \forall k \in [N],
				\sum_{i=1}^{N}q_{i} = \ q_{sum}.$  
		\end{problem}
	\end{mdframed}
	
	It can be noted that the expressions in \eqref{eq:pdp_expression_SF} and \eqref{eq:pdp_expression_FF} look similar. However, due to the presence of the distinct order Marcum-Q functions, the strategies for solving the optimization problems are different.  		 
	\subsection{Analysis on the Near-Optimal ARQ Distribution for CC-HARQ-FF Based Non-Cumulative Network}
	\label{sec:necc_and_suff_FF}
	Before we obtain the necessary and sufficient conditions on the near-optimal ARQ distribution, we decide which link should get more ARQs than others. 
	\begin{theorem}
		\label{LOShighlow_FF}
		For a given vector $\mathbf{c}$ and a high SNR regime in a CC-HARQ-FF based multi-hop network, the solution to Problem \ref{approx_opt_problem_2_ff} satisfies the property that whenever $c_{i} \geq c_{j}$, we have $q_{i} \leq q_{j} \ \forall i,j \ \in [N]$. 
	\end{theorem}
	\begin{IEEEproof}
	Along the similar lines of Theorem \ref{LOShighlow_SF}, we only consider the $i$-th and $j$-th links and assume that they have $`q'$ ARQs to start with, and $c_{i} \geq c_{j}$. Let $A=\tilde{P}^{f}_{iq}+ \tilde{P}^{f}_{j(q+1)}(1-\tilde{P}^{f}_{iq})$ and $B=\tilde{P}^{f}_{i(q+1)}+\tilde{P}^{f}_{jq}(1-\tilde{P}^{f}_{i(q+1)}) $ with $q_{i}=q_{j}=q$. By rearranging the terms in $A$ and $B$, we get $A=\tilde{P}^{f}_{iq}+ \tilde{P}^{f}_{j(q+1)}-\tilde{P}^{f}_{iq}\tilde{P}^{f}_{j(q+1)}$ and $B=\tilde{P}^{f}_{i(q+1)}+\tilde{P}^{f}_{jq}-\tilde{P}^{f}_{i(q+1)}\tilde{P}^{f}_{jq}$. On taking the difference between $A$ and $B$, we get $A-B=\tilde{P}^{f}_{iq}-\tilde{P}^{f}_{i(q+1)}+ \tilde{P}^{f}_{j(q+1)}-\tilde{P}^{f}_{jq}-\tilde{P}^{f}_{iq}\tilde{P}^{f}_{j(q+1)}+\tilde{P}^{f}_{i(q+1)}\tilde{P}^{f}_{jq}$. Using the approximation on $\tilde{P}_{k}^{f}$ in  \eqref{eq:approximation_P_k_ff}, we can rewrite the difference as 
	\begin{small}
		\begin{eqnarray*}
			A-B &=& \underbrace{\frac{1}{q!}\bigg(\frac{b_{i}^{2}}{2} e^{\frac{-a_{i}^{2}}{2}}\bigg)^{q}\bigg(1-\frac{1}{(q+1)}\bigg(\frac{b_{i}^2}{2}\bigg)e^{\frac{-a_{i}^{2}}{2}}\bigg)} - \underbrace{\frac{1}{q!}\bigg(\frac{b_{j}^{2}}{2}e^{\frac{-a_{j}^{2}}{2}}\bigg)^{q}\bigg(1-\frac{1}{(q+1)}\bigg(\frac{b_{j}^{2}}{2}\bigg)e^{\frac{-a_{j}^{2}}{2}}\bigg)} \\ & &
			+\underbrace{\frac{1}{(q!)^{2}}\bigg(\frac{b_{i}^{2}b_{j}^{2}}{2}\bigg)^{q}\frac{e^{(-a_{i}^{2}-a_{j}^{2})/2}}{q+1}\bigg(-\frac{b_{j}^{2}}{2}e^{\frac{-a_{j}^{2}}{2}}+\frac{b_{i}^{2}}{2}e^{\frac{-a_{i}^{2}}{2}}\bigg)}.   
		\end{eqnarray*}
	\end{small}
	From the above equation, it can be observed that the first term in the above equation is positive, and the second term is negative. Also, at high SNR, we can ignore high power terms of $b_{i} \ \text{or} \ b_{j}$. Therefore, by neglecting the third term from the above equation, we get 
	\begin{small}
		\begin{eqnarray*}
			A-B &=& \underbrace{\frac{1}{q!}\bigg(\frac{b_{i}^{2}}{2}e^{\frac{-a_{i}^{2}}{2}}\bigg)^{q}\bigg(1-\frac{1}{(q+1)}\bigg(\frac{b_{i}^2}{2}\bigg)e^{\frac{-a_{i}^{2}}{2}}\bigg)} - \underbrace{\frac{1}{q!}\bigg(\frac{b_{j}^{2}}{2}e^{\frac{-a_{j}^{2}}{2}}\bigg)^{q}\bigg(1-\frac{1}{(q+1)}\bigg(\frac{b_{j}^{2}}{2}\bigg)e^{\frac{-a_{j}^{2}}{2}}\bigg)}. 
		\end{eqnarray*}
	\end{small}
	Now, we want to estimate whether the first term is greater than the second term or not. Let us start by assuming that the second term is greater than the first term. Therefore, we can write
	\begin{small}
		\begin{eqnarray*}  \frac{1}{q!}\bigg(\frac{b_{j}^{2}}{2}e^{\frac{-a_{j}^{2}}{2}}\bigg)^{q}\bigg(1-\frac{1}{(q+1)}\bigg(\frac{b_{j}^{2}}{2}\bigg)e^{\frac{-a_{j}^{2}}{2}}\bigg) \geq \frac{1}{q!}\bigg(\frac{b_{i}^{2}}{2}e^{\frac{-a_{i}^{2}}{2}}\bigg)^{q}\bigg(1-\frac{1}{(q+1)}\bigg(\frac{b_{i}^2}{2}\bigg)e^{\frac{-a_{i}^{2}}{2}}\bigg). 
		\end{eqnarray*} 
	\end{small}
	After rearranging and solving the above equation, we get 
		\begin{equation*} \bigg(\frac{e^{-a_{j}^{2}/2}}{e^{-a_{i}^{2}/2}}\bigg)^{q}\Bigg(\frac{1-\frac{b_{j}^{2}e^{-a_{j}^{2}/2}}{2(q+1)}}{1-\frac{b_{i}^{2}e^{-a_{i}^{2}/2}}{2(q+1)}}\Bigg) \geq \bigg(\frac{b_{i}^{2}}{b_{j}^{2}}\bigg)^{q}.
		\end{equation*}
	Furthermore, by expanding the $b_{k}^{2} = \frac{2\phi}{(1-c_{k})} \forall k \in \{i,j\}$ on RHS of the equation and making $\frac{b_{j}^{2}e^{-a_{j}^{2}/2}}{2(q+1)} \approx 0$ and $\frac{b_{i}^{2}e^{-a_{i}^{2}/2}}{2(q+1)} \approx 0$, we obtain $\Big(\frac{e^{-a_{j}^{2}/2}}{e^{-a_{i}^{2}/2}}\Big)^{q} \geq \big(\frac{1-c_{j}}{1-c_{i}}\big)^{q}$. By taking the logarithm on both the sides, we get
	\begin{eqnarray} 
		\label{eq:trueOrfalse}	
		q\bigg(\frac{c_{i}-c_{j}}{(1-c_{j})(1-c_{i})}\bigg) \geq q\log \bigg(\frac{1-c_{j}}{1-c_{i}}\bigg).
	\end{eqnarray} 
	Using the standard inequality $\log x \leq x-1$, observe that \eqref{eq:trueOrfalse} is true for the given condition i.e. $c_{i} \geq c_{j}$. It implies that $A-B \leq 0$ and higher LOS must be given lower ARQs. 	  
	\end{IEEEproof}
	
	Similar to Theorem \ref{NeccSuff}, we can first define a local minima for the CC-HARQ-FF strategy (similar to that of Definition \ref{def:lm}, however by replacing $\tilde{p}^{s}_{d}$ by $\tilde{p}^{f}_{d}$), and then provide the necessary and sufficient conditions on near-optimal ARQ distribution for CC-HARQ-FF.	
		\begin{theorem}
			\label{thm:NeccSuff_FF}
		For a given $N$-hop network with LOS vector $\mathbf{c}$, the ARQ distribution $\mathbf{q}^{*} = [q^{*}_{1}, q^{*}_{2}, \ldots, q^{*}_{N}]$ is said to be a local minima for CC-HARQ-FF strategy if and only if $q^{*}_{i}$ and $q^{*}_{j}$ for $i \neq j$ satisfy
		\begin{small}
		\begin{eqnarray} 
			\label{Necc_Suff_FF_1}
			& & \frac{B_{i}^{q^{*}_{i}}}{q^{*}_{i}!} \bigg(1- \frac{B_{i}}{q_{i}^{*}+1}\bigg) + \frac{B_{j}^{(q^{*}_{j}-1)}}{(q^{*}_{j}-1)!} \bigg(\frac{B_{j}}{q_{j}^{*}}-1\bigg) \leq 
			\frac{B_{i}^{q^{*}_{i}}B_{j}^{(q^{*}_{j}-1)}}{q^{*}_{i}!(q^{*}_{j}-1)!}  \bigg(\frac{B_{j}}{q_{j}^{*}}- \frac{B_{i}}{q_{i}^{*}+1}\bigg), \\
			\label{Necc_Suff_FF_2}
			& & \frac{B_{i}^{(q^{*}_{i}-1)}}{(q^{*}_{i}-1)!} \bigg( 1- \frac{B_{i}}{q_{i}^{*}} \bigg) + \frac{B_{j}^{q^{*}_{j}}}{q^{*}_{j}!} \bigg(\frac{B_{j}}{q_{j}^{*}+1}-1\bigg) \geq 
			\frac{B_{i}^{(q^{*}_{i}-1)}B_{j}^{q^{*}_{j}}}{(q^{*}_{i}-1)!q^{*}_{j}!}   \bigg( \frac{B_{j}}{q_{j}^{*}+1}-\frac{B_{i}}{q_{i}^{*}}\bigg),  
		\end{eqnarray}
	\end{small}
		where $B_{i}= (\phi e^{-a_{i}^{2}/2})/(1-c_{i})$ and $B_{j}= (\phi e^{-a_{j}^{2}/2})/(1-c_{j})$ with $\phi = (2^{R}-1)/\gamma$. 
	\end{theorem}
	\begin{IEEEproof}
	By using the definition of local minima, we have the inequalities $\tilde{p}_{d}^{f}(\mathbf{q}^{*}) \leq \tilde{p}_{d}^{f}(\hat{\mathbf{q}}_{+}), \mbox{ and } \tilde{p}_{d}^{f}(\mathbf{q}^{*}) \leq \tilde{p}_{d}^{f}(\hat{\mathbf{q}}_{-}),$ where $\tilde{p}_{d}^{f}(\mathbf{q}^{*})$, $\tilde{p}_{d}^{f}(\hat{\mathbf{q}}_{+})$ and $\tilde{p}_{d}^{f}(\hat{\mathbf{q}}_{-})$ represent the PDP evaluated at the distributions $\mathbf{q}^{*}$, $\hat{\mathbf{q}}_{+}$, and $\hat{\mathbf{q}}_{-}$, respectively. Because $\hat{\mathbf{q}}_{+}$ and $\hat{\mathbf{q}}_{-}$ differ only in the last two positions and the structure of the PDP, the above inequalities are equivalent to
\begin{small} 
	\begin{eqnarray*} 
		\tilde{P}^{f}_{iq^{*}_{i}} +\tilde{P}^{f}_{jq^{*}_{j}}(1-\tilde{P}^{f}_{iq^{*}_{i}}) \leq \tilde{P}^{f}_{i(q^{*}_{i}+1)} + \tilde{P}^{f}_{j(q^{*}_{j}-1)}(1-\tilde{P}^{f}_{i(q^{*}_{i}+1)}); 
		\tilde{P}^{f}_{iq^{*}_{i}} +\tilde{P}^{f}_{jq^{*}_{j}}(1-\tilde{P}^{f}_{iq^{*}_{i}}) \leq \tilde{P}^{f}_{i(q^{*}_{i}-1)} + \tilde{P}^{f}_{j(q^{*}_{j}+1)}(1-\tilde{P}^{f}_{i(q^{*}_{i}-1)}).
	\end{eqnarray*} 
\end{small}
Now, by using the approximation on $\tilde{P}^{f}_{iq}$ from \eqref{eq:approximation_P_k_ff}, we can rewrite the above two inequalities as 	
\begin{small}
	\begin{eqnarray*} 
		& & \frac{1}{q^{*}_{i}!} \bigg(\frac{\phi e^{\frac{-a_{i}^{2}}{2}}}{1-c_{i}}\bigg)^{q_{i}^{*}}  + \frac{1}{q^{*}_{j}!} \bigg(\frac{\phi e^{\frac{-a_{j}^{2}}{2}}}{1-c_{j}}\bigg)^{q_{j}^{*}}  \bigg[1 - \frac{1}{q^{*}_{i}!} \bigg(\frac{\phi e^{\frac{-a_{i}^{2}}{2}}}{1-c_{i}}\bigg)^{q_{i}^{*}} \bigg]  \leq  \frac{1}{(q^{*}_{i}+1)!} \\ 
		& & \bigg(\frac{\phi e^{\frac{-a_{i}^{2}}{2}}}{1-c_{i}}\bigg)^{(q^{*}_{i}+1)}  + \frac{1}{(q_{j}^{*}-1)!} \bigg(\frac{\phi e^{\frac{-a_{j}^{2}}{2}}}{1-c_{j}}\bigg)^{(q_{j}^{*}-1)} \bigg[1 - \frac{1}{(q^{*}_{i}+1)!} \bigg(\frac{\phi e^{\frac{-a_{i}^{2}}{2}}}{1-c_{i}}\bigg)^{(q^{*}_{i}+1)} \bigg], \\
		& & \frac{1}{q^{*}_{i}!} \bigg(\frac{\phi e^{\frac{-a_{i}^{2}}{2}}}{1-c_{i}}\bigg)^{q_{i}^{*}}  + \frac{1}{q^{*}_{j}!} \bigg(\frac{\phi e^{\frac{-a_{j}^{2}}{2}}}{1-c_{j}}\bigg)^{q_{j}^{*}} \bigg[1 - \frac{1}{q^{*}_{i}!} \bigg(\frac{\phi e^{\frac{-a_{i}^{2}}{2}}}{1-c_{i}}\bigg)^{q_{i}^{*}} \bigg]  \leq  \frac{1}{(q^{*}_{i}-1)!} \\ 
		& & \bigg(\frac{\phi e^{\frac{-a_{i}^{2}}{2}}}{1-c_{i}}\bigg)^{(q^{*}_{i}-1)}  + \frac{1}{(q_{j}^{*}+1)!} \bigg(\frac{\phi e^{\frac{-a_{j}^{2}}{2}}}{1-c_{j}}\bigg)^{(q_{j}^{*}+1)} \bigg[1 - \frac{1}{(q^{*}_{i}-1)!} \bigg(\frac{\phi e^{\frac{-a_{i}^{2}}{2}}}{1-c_{i}}\bigg)^{(q^{*}_{i}-1)} \bigg].
	\end{eqnarray*} 
\end{small}
Assuming $B_{i}=(\phi e^{\frac{-a_{i}^{2}}{2}})/(1-c_{i})$ for $i \in [N]$, we can rewrite the above two inequalities as 
\begin{small}
	\begin{eqnarray*} 
		& & \frac{1}{q^{*}_{i}!} B_{i}^{q_{i}^{*}}  + \frac{1}{q^{*}_{j}!} B_{j}^{q^{*}_{j}}  \bigg(1 - \frac{1}{q^{*}_{i}!} B_{i}^{q_{i}^{*}} \bigg)  \leq  \frac{1}{(q^{*}_{i}+1)!}  B_{i}^{(q^{*}_{i}+1)}  + \frac{1}{(q_{j}^{*}-1)!} B_{j}^{(q_{j}^{*}-1)} 
		\bigg(1 - \frac{1}{(q^{*}_{i}+1)!} B_{i}^{(q^{*}_{i}+1)} \bigg), \\
		& & \frac{1}{q^{*}_{i}!} B_{i}^{q_{i}^{*}}  + \frac{1}{q^{*}_{j}!} B_{j}^{q^{*}_{j}}  \bigg(1 - \frac{1}{q^{*}_{i}!} B_{i}^{q_{i}^{*}} \bigg)   \leq  \frac{1}{(q^{*}_{i}-1)!}  B_{i}^{(q^{*}_{i}-1)}  + \frac{1}{(q_{j}^{*}+1)!} B_{j}^{(q_{j}^{*}+1)} \bigg(1 - \frac{1}{(q^{*}_{i}-1)!} B_{i}^{(q^{*}_{i}-1)} \bigg),
	\end{eqnarray*} 
\end{small}
respectively. By rearranging the above two inequalities, we can obtain the two claimed inequalities. 
	\end{IEEEproof}
   \subsection{Low-Complexity Algorithm for CC-HARQ-FF based Non-Cumulative Network}
	It can be observed from \eqref{Necc_Suff_FF_1} and \eqref{Necc_Suff_FF_2} that the conditions are non-linear with integer constraints. Due to higher-order non-linearity, it is not possible to solve the conditions analytically on the near-optimal ARQ distribution. Hence, towards finding the near-optimal ARQ distribution, in this section, first, we approximate the necessary and sufficient conditions at high SNR. After that, we propose a Fold-To-Make-List (FTML) low-complexity algorithm based on numerical-methods, as given in Algorithm \ref{List_algoritm_FF}. 
	
	Towards approximating the necessary and sufficient conditions at high SNR, let us rearrange \eqref{Necc_Suff_FF_2} as 
	\begin{eqnarray*} 
		\frac{B_{i}^{(q^{*}_{i}-1)}}{(q^{*}_{i}-1)!} \bigg(\frac{B_{i}}{q_{i}^{*}} - 1 \bigg) + \frac{B_{j}^{q^{*}_{j}}}{q^{*}_{j}!} \bigg(1- \frac{B_{j}}{q_{j}^{*}+1}\bigg) \leq 
		\frac{B_{i}^{(q^{*}_{i}-1)}B_{j}^{q^{*}_{j}}}{(q^{*}_{i}-1)!q^{*}_{j}!}   \bigg( \frac{B_{i}}{q_{i}^{*}}-\frac{B_{j}}{q_{j}^{*}+1}\bigg),  
	\end{eqnarray*}
	and add the above inequality with \eqref{Necc_Suff_FF_1}, to obtain 
	\begin{scriptsize} 
	\begin{eqnarray*} 
		\frac{B_{i}^{(q^{*}_{i}-1)}}{(q^{*}_{i}-1)!} \bigg(\frac{2B_{i}}{q_{i}^{*}}-1- \frac{B_{i}^{2}}{q_{i}^{*}(q_{i}^{*}+1)}\bigg) + \frac{B_{j}^{(q^{*}_{j}-1)}}{(q^{*}_{j}-1)!} \bigg(\frac{2B_{j}}{q_{j}^{*}}-1- \frac{B_{j}^{2}}{q_{j}^{*}(q_{j}^{*}+1)}\bigg) \leq 
		\frac{B_{i}^{(q^{*}_{i}-1)}B_{j}^{(q^{*}_{j}-1)}}{(q^{*}_{i}-1)!(q^{*}_{j}-1)!} \bigg(\frac{2B_{i}B_{j}}{q_{i}^{*}q_{j}^{*}}- \frac{B_{i}^{2}}{q_{i}^{*}(q_{i}^{*}+1)}-\frac{B_{j}^{2}}{q_{j}^{*}(q_{j}^{*}+1)}\bigg). 
	\end{eqnarray*}
\end{scriptsize}At high SNR, first, we focus on the LHS of the expression, wherein we can ignore the terms which contain $B_{i}^{2}$ and $B_{j}^{2}$. This is because they are negligible compared to unity. Likewise, on the RHS of the expression, we can ignore the terms which contain $B_{i}^{2}$, $B_{j}^{2}$ and $B_{i}B_{j}$ due to the same reason as specified above. Therefore, by ignoring the above-mentioned terms, we can obtain the inequalities as shown below 
$\vspace{-0.9cm}$
	\begin{eqnarray*} 
		\frac{B_{i}^{(q^{*}_{i}-1)}}{(q^{*}_{i}-1)!} \bigg(\frac{2B_{i}}{q_{i}^{*}}-1\bigg) + \frac{B_{j}^{(q^{*}_{j}-1)}}{(q^{*}_{j}-1)!} \bigg(\frac{2B_{j}}{q_{j}^{*}}-1 \bigg) & \leq & 0, \\ 
		\frac{B_{i}^{q^{*}_{i}}}{q^{*}_{i}!} \bigg(2-\frac{q_{i}^{*}}{B_{i}}\bigg) + \frac{B_{j}^{q^{*}_{j}}}{q^{*}_{j}!} \bigg(2-\frac{q_{j}^{*}}{B_{j}} \bigg) & \leq & 0,
	\end{eqnarray*}
\noindent where the second inequality can be obtained from the first inequality using algebraic manipulations. Furthermore, there are multiple ways in which $q_{i}^{*}$ and $q_{j}^{*}$ can follow the above inequality. One such way is to force that $\frac{B_{i}^{q^{*}_{i}}}{q^{*}_{i}!}= \frac{B_{j}^{q^{*}_{j}}}{q^{*}_{j}!}$. This is because, by writing $ \frac{B_{j}^{q^{*}_{j}}}{q^{*}_{j}!} = \frac{B_{i}^{q^{*}_{i}}}{q^{*}_{i}!}$, we can obtain $\frac{B_{i}^{q^{*}_{i}}}{q^{*}_{i}} \Big( 4 - \frac{q_{i}^{*}}{B_{i}}-\frac{q_{j}^{*}}{B_{j}}\Big) \leq 0,$ which is also true at high SNR values. Therefore, we impose the condition $\frac{B_{i}^{q^{*}_{i}}}{q^{*}_{i}!} = \frac{B_{j}^{q^{*}_{j}}}{q^{*}_{j}!}$ for every $i,j \in [N]$ using the approximation on the sufficient conditions. Furthermore, by using the Stirling's formulae i.e. $n ! \approx \sqrt{2 \pi n}(n/e)^{n}$, and the further solving the equations, we get 
	\begin{eqnarray}
		\label{non_linear_FF}	
		\sqrt{ q^{*}_{j}} \bigg(\frac{q^{*}_{j}}{e B_{j}}\bigg)^{q^{*}_{j}} & = & \sqrt{q^{*}_{i}}\bigg(\frac{q^{*}_{i}}{e B_{i}}\bigg)^{q^{*}_{i}}. 
	\end{eqnarray} 
	Now, to obtain the near-optimal ARQ distribution by using the above expression, we formulate an equivalent optimization problem of Problem \ref{approx_opt_problem_2_ff} as given below:
	\vspace{0.5cm}
	\begin{mdframed}
		\begin{problem}
			\label{opt_problem_ff_approx}
			For a given $N$, $\mathbf{B}=\{B_{1}, B_{2}, \ldots, B_{N} \}$, and $q_{sum}$, find $\mathbf{q} = [q^{*}_{1}, q^{*}_{2}, \ldots, q^{*}_{N}]$ such that $$\sqrt{ q^{*}_{j}} \bigg(\frac{q^{*}_{j}}{e B_{j}}\bigg)^{q^{*}_{j}} = \sqrt{ q^{*}_{i}}\bigg(\frac{q^{*}_{i}}{e B_{i}}\bigg)^{q^{*}_{i}}, ~\forall i, j \in [N] \ \mbox{where} \ i \neq j, q^{*}_{i} \geq 1, q^{*}_{i}\in \mathbb{Z_{+}}, ~\forall i \in [N], \ \sum_{i=1}^{N}q^{*}_{i} = \ q_{sum}.$$  
		\end{problem}
	\end{mdframed}	
	
	\begin{algorithm}
		\caption{\label{List_algoritm_FF}Fold-To-Make-List (FTML)}
		\label{LIST_algorithm_FF}
		\begin{algorithmic}[1]
	     \begin{small}
			\Require $\mathbf{B}$, $q_{sum}$, $\mathbf{c} = [c_{1}, c_{2}, \ldots, c_{N}]$.
			\Ensure $\mathcal{L}^{f} \subset \mathbb{S}$ - List of ARQ distributions in search space $\mathbb{S}$.
			\State $\mathcal{L}^{I} = \{\phi\}$ - Internal array of list of ARQ distributions in $\mathbb{Z}_{+}$.
			\For {$q_{1}^{*}=1:q_{sum}-(N-1)$}
			\State $\mathcal{L}^{I}_{q_{1}^{*},1} =\{ q_{1}^{*} \}$.
			\For {$j = 2:N$}
			\State Assign $m_{1} = 1$, $m_{2} = q_{sum}-(N-1)$.
			\While{$m_{2}-m_{1} \geq 2$}
			\For{$k = 1:2$}
			\State Calculate $\mathfrak{D_{j}}(m_{k}) =  \sqrt{m_{k}}(m_{k}/e B_{j})^{m_{k}}$; Compute $d_{k}=  |\mathfrak{D_{j}}(m_{k})-\mathfrak{C_{1}}(q_{1}^{*})|$.
			\EndFor
			\If{$d_{1} \geq d_{2}$}
			\State Assign $m_{1} = \floor{\frac{m_{1}+m_{2}}{2}}$.
			\Else
			\State Assign $m_{2} = \floor{\frac{m_{1}+m_{2}}{2}}$.
			\EndIf
			\EndWhile
			\For{$l = 1:2$}
			\State Compute $\mathfrak{D_{j}}(m_{l}) =  \sqrt{m_{l}}(m_{l}/e B_{j})^{m_{l}}$; Compute $e_{l}=  |\mathfrak{D_{j}}(m_{l})-\mathfrak{C_{1}}(q_{1}^{*})|$.
			\EndFor
			\If{$e_{1} \leq e_{2}$}
			\State Assign $q_{j}^{*} = m_{1}$.
			\Else
			\State Assign $q_{j}^{*} = m_{2}$.
			\EndIf
			\State Insert $q_{j}^{*}$ in $\mathcal{L}^{I}_{q_{1}^{*},j}$. 
			\EndFor
			\EndFor
			\For {$t= 1 : \text{rows}(\mathcal{L}^{I})$}
			\State Compute $I_{t} = \left(\sum_{i = 1}^{N} \mathcal{L}^{I}_{t,i}\right) - q_{sum}$; $\mathcal{L}^{f}_{t} = \{\mathbf{q} \in \mathbb{S} ~|~ d(\mathbf{q}, \tilde{\mathbf{q}}) = I_{t}, q_{i}, q_{j} \geq 1\}$.
			\EndFor 
			\end{small}
		\end{algorithmic}
	\end{algorithm}
	In the rest of this section, our objective is to find the near-optimal ARQ distribution for Problem \ref{opt_problem_ff_approx}. Note that the equation in Problem \ref{opt_problem_ff_approx} is non-tractable because we cannot solve it analytically due to the non-linearity and integer constraints. Therefore, we obtain near-optimal ARQ distribution with the help of an FTML low-complexity algorithm based on numerical-methods as proposed in Algorithm \ref{LIST_algorithm_FF}. In Algorithm \ref{LIST_algorithm_FF}, we define $\mathfrak{D_{j}} (q_{j}^{*}) \triangleq \sqrt{ q^{*}_{j}}(q^{*}_{j}/e B_{j})^{q^{*}_{j}}$ and  $\mathfrak{C_{i}}(q_{i}^{*})  \triangleq \sqrt{ q^{*}_{i}}(q^{*}_{i}/e B_{i})^{q^{*}_{i}} \ \forall i,j \in [N] \ \& \ i \neq j$, where $\mathfrak{C_{i}}(q_{i}^{*})$ is a constant term for a given $q_{i}^{*}$, and thus we can rewrite \eqref{non_linear_FF} as $\mathfrak{D_{j}} (q_{j}^{*}) = \mathfrak{C_{i}}(q_{i}^{*})$.  
	
	Towards explaining Algorithm \ref{List_algoritm_FF}, we observe that for a fixed $q_{i}^{*}$, the RHS of  $\mathfrak{D_{j}} (q_{j}^{*}) = \mathfrak{C_{i}}(q_{i}^{*})$ is a constant term. Therefore, by fixing $q_{1}^{*}$ (wherein $q_{1} \in [q_{sum}-(N-1)]$), we can find the ARQ distribution $q_{j}^{*}$, $\forall j \in [2,N] $, by forming pairs such as $(q_{1}^{*},q_{2}^{*}),(q_{1}^{*},q_{3}^{*}),\ldots$ , and so on. Thereby, for each possible value of $q_{1}^{*}$, the other ARQ distribution can be calculated. Thus, a list of suitable ARQ distributions can be created, which is denoted by $\mathcal{L}^{f}$ (in Algorithm \ref{List_algoritm_FF}) and from the list, the near-optimal ARQ distribution can be obtained. Furthermore, as the LHS of $\mathfrak{D_{j}} (q_{j}^{*}) = \mathfrak{C_{i}}(q_{1}^{*})$ is a non-linear term, we can apply a folding technique based on numerical-methods to solve it. In particular, we use iterations, wherein the search space for $q_{j}^{*}$ is reduced by half in each iteration such that it converges into an appropriate value of $q_{j}^{*}$. We validate the accuracy of our Algorithm \ref{List_algoritm_FF} using simulations in the next section.
	\subsection{Simulation Results for CC-HARQ-FF based Non-Cumulative Network}
\begin{figure*}[h]
		\centering \includegraphics[scale=0.42]{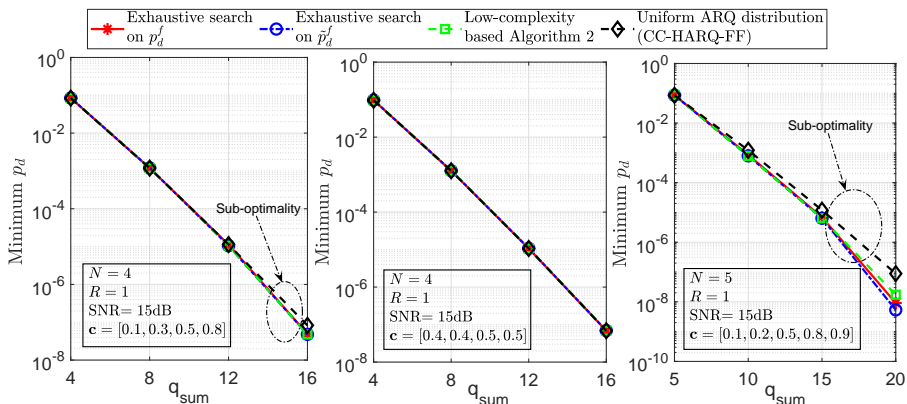}
		\vspace{-0.5cm}	
		\centering{\caption{PDP comparison in CC-HARQ-FF based non-cumulative network between (i) exhaustive search using original PDP expression, $p_{d}^{f}$, (ii) exhaustive search using approximated PDP expression, $\tilde{p}_{d}^{f}$, (iii) proposed FTML low-complexity algorithm (given in Algorithm \ref{LIST_algorithm_FF}), and (iv) uniform ARQ distribution.}
				\label{fig:PDP_comparison_HARQ}}
	\end{figure*}

	To validate all the results on the optimal ARQ distribution, we present the plots in Fig. \ref{fig:PDP_comparison_HARQ} for $N=4$ and $N=5$ at different $q_{sum}$. The following conclusions can be drawn through the plots: (i) the minimum PDP obtained by using an approximated expression in \eqref{eq:Approx_pdp_expression_ff} almost coincides with the minimum PDP obtained from the original PDP expression in \eqref{eq:pdp_expression_FF}. Hence, it shows the accuracy of our approximation on the Marcum-Q function of a higher-order, (ii) the proposed low-complexity algorithm gives us the near-optimal ARQ distribution, hence validating its accuracy, and (iii) near-optimal ARQ distribution that we obtain from our algorithm is close to the uniform ARQ distribution (unlike the case of CC-HARQ-SF). We believe that this behaviour is attributed to the fact that the dominating factor in the packet drop probability at the $i$-th hop is $q_{i}!$. This way, while solving Problem \ref{opt_problem_ff_approx} (which is equivalent to solving the Problem \ref{approx_opt_problem_2_ff}) using numerical-methods, the near-optimal ARQ distribution is closer to the uniform distribution. 
\vspace{-0.2cm}
	\begin{figure}[h!]
		\centering \includegraphics[scale=0.42]{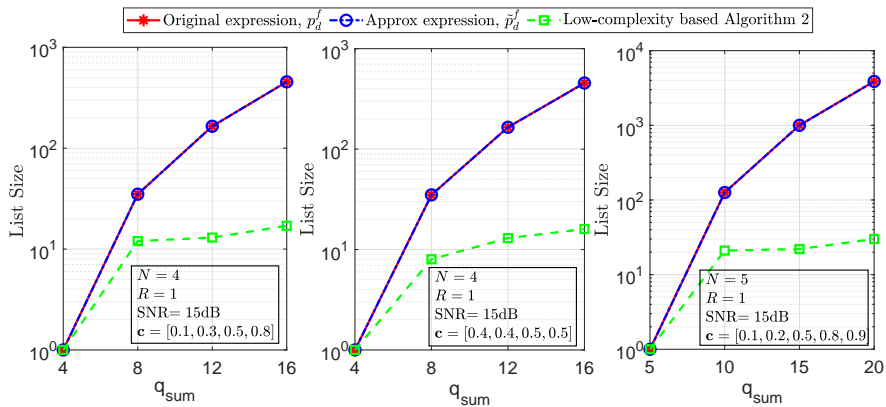}
		\vspace{-0.5cm}	
		\centering{\caption{Comparison of list size between (i) exhaustive search in original PDP expression, $p_{d}^{f}$ (in \eqref{eq:pdp_expression_FF}), (ii) exhaustive search in approximated PDP expression, $\tilde{p}_{d}^{f}$ (in \eqref{eq:Approx_pdp_expression_ff}), and (iii) the proposed FTML low-complexity algorithm for CC-HARQ-FF based non-cumulative network.}
				\label{fig:ListSize_FF}}
	\end{figure}Furthermore, similar to Section \ref{sec:Sims}, the list size of the exhaustive search in case of original expression (in \eqref{eq:pdp_expression_FF}) and the approximated expression (in \eqref{eq:Approx_pdp_expression_ff}) is ${q_{sum}-1}\choose{N-1}$. However, by using the proposed low-complexity method given in Algorithm \ref{LIST_algorithm_FF}, we can reduce the list size significantly as shown in Fig. \ref{fig:ListSize_FF}. In the next section, we provide results for the CC-HARQ-FF based fully-cumulative network. 
	
	\section{CC-HARQ-FF Based Fully-Cumulative Network}
	\label{sec:fully_cumm_FF}
	The idea of the fully-cumulative network is similar to that in Section \ref{sec:Sims_delay_analysis_SF} with the only exception that the PDP expression, in this case, will consider fast fading intermediate channels. In the following theorem, we present our results on the optimal ARQ distribution. 
	\begin{figure}[h!]
	\centering \includegraphics[scale = 0.45]{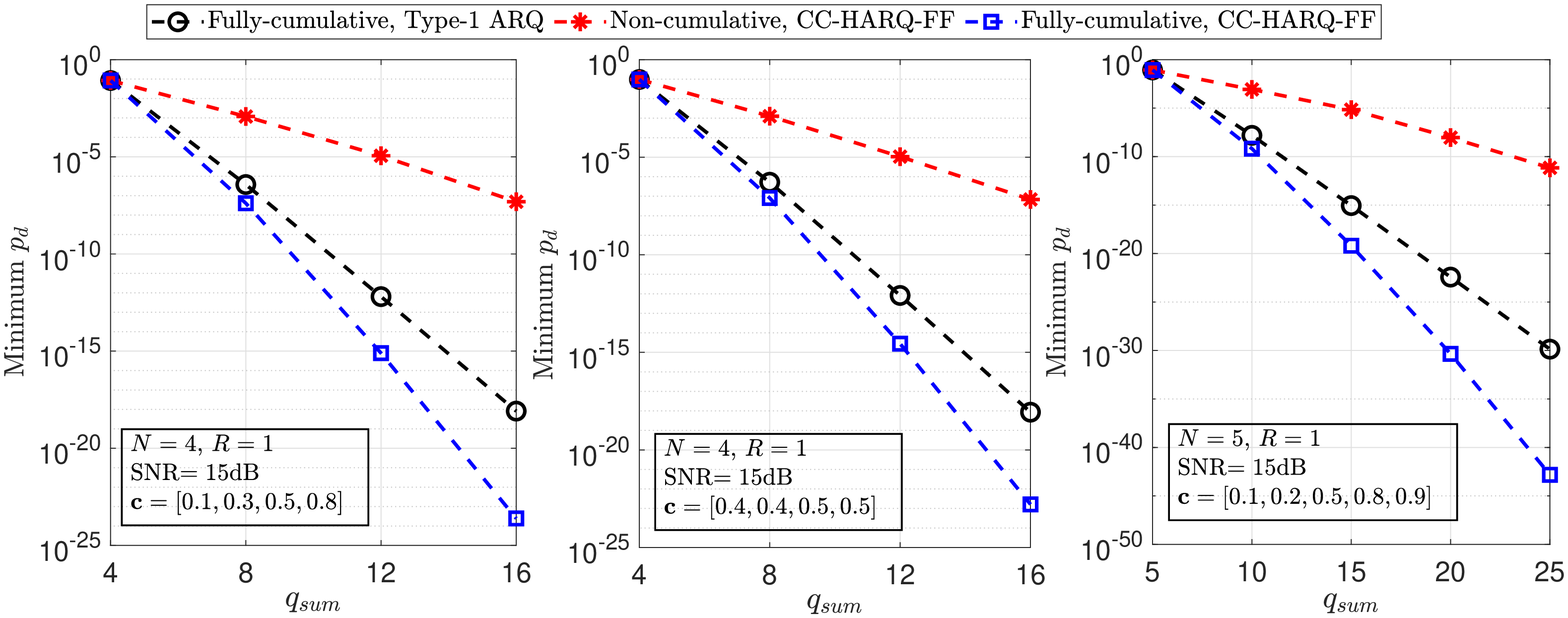}
	\vspace{-0.5cm}
	\centering{\caption{Comparison of PDP for the CC-HARQ-FF based non-cumulative network , CC-HARQ-FF based fully-cumulative network, and Type-1 ARQ based fully-cumulative network at various values of $q_{sum}$}. 
	\label{fig:fully_cumm_FF}}
	\end{figure}
	\begin{theorem}
		\label{Fully_cumm_fastfading}
		For a given ARQ distribution $\mathbf{q} = [q_{1}, q_{2}, \ldots, q_{N}]$ and $q_{sum}$ in a CC-HARQ-FF based fully-cumulative network, the optimal ARQ distribution can be given by $[q_{sum}, 0, \ldots, 0]$.
	\end{theorem}
	\begin{IEEEproof}
	We omit the proof as it can be proved along the similar lines of the proof for Theorem \ref{thm:Fully_cumm_slowfading}. However, in this proof, we need to invoke the definition $P_{kq_{k}}^{f} = 1$ when $q_{k} = 0$, for $k \in [N]$. 
	\end{IEEEproof}

	To showcase the benefit of fully-cumulative network, we plot the results for a $4$-hop network and a $5$-hop network in Fig. \ref{fig:fully_cumm_FF}, where we observe a significant reduction in the PDP when we use the fully-cumulative scheme over a non-cooperative scheme. Also, it can be observed that CC-HARQ-FF based fully-cumulative strategy performs better than the Type-1 ARQ based fully-cumulative strategy \cite{our_work_TWC_1}, which represents the benefit of using CC-HARQ-FF strategy over Type-1 ARQ strategy. However, as mentioned in Section \ref{sec:fully_cumm_SF}, the fully-cumulative network uses a counter that needs to be updated at each hop to convey the residual ARQs to the next node in the chain. Therefore, it may increase the delay overhead compared to the non-cumulative network. This is the cost associated with the fully-cumulative network in order to reduce the PDP from that of a non-cumulative network. 
	
	\section{Simulation results on delay analysis for CC-HARQ-FF Strategy}
	\label{sec:Sims_delay_analysis_FF}
\begin{figure}[h!]
		\centering \includegraphics[scale=0.45]{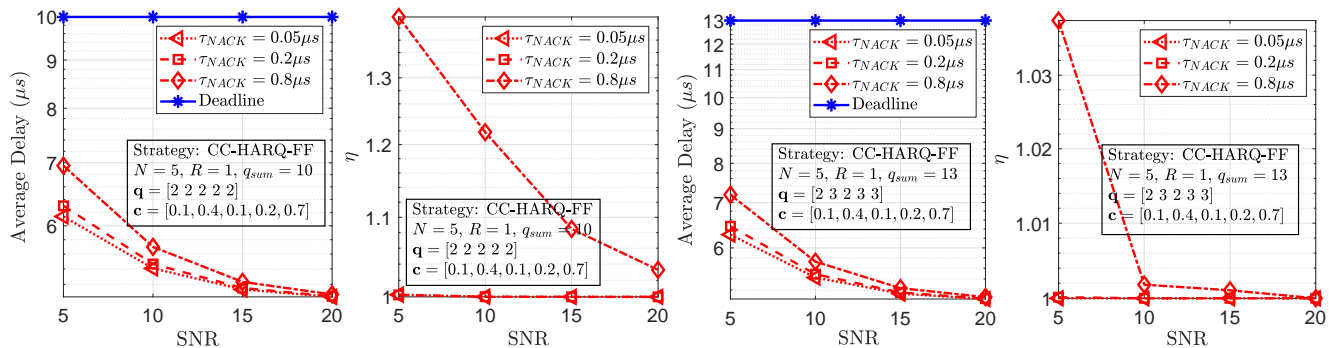}
		\vspace{-0.51cm}
		\centering{\caption{Variation of average delay on the packets and the deadline violation parameter ($\eta$) for various $\tau_{NACK}$ in CC-HARQ-FF strategies.  
				\label{fig:average_delay_FF}}}
	\end{figure}

	Similar to Section \ref{sec:Sims_delay_analysis_SF}, in this section, we present a thorough analysis on the end-to-end delay for CC-HARQ-FF strategies by using the same set of delay and deadline-violation metrics. 
	First, we show that the packets in a non-cumulative network reach the destination before the deadline with a higher probability, provided that the delay overheads from ACK/NACK are sufficiently small. To demonstrate the results, we obtain $q_{sum}$ as $\floor{\frac{\tau_{total}}{\tau_{p}+\tau_{d}}}$ without considering the resources for ACK/NACK in the reverse channel, where $\tau_{total}, \tau_{d}$ and $\tau_{p}$ are as defined in Section \ref{sec:system_model}. Subsequently, we introduce a different resolution of delays from NACK, say $\tau_{NACK}$ time units, and then observe the impact on the end-to-end delay on the packets. Assuming $\tau_{p} + \tau_{d} = 1 ~\mu s$, we set the deadline for end-to-end packet delay as $q_{sum} ~\mu s$. Then, by sending an ensemble of $10^6 $ packets to the destination through the CC-HARQ-FF strategy, we compute the following metrics when $\tau_{NACK} \in \{0.05, 0.2, 0.8\}$ in $\mu s$: (i) $P_{Drop}$, (ii) $P_{Deadline}$, and finally, (iii) the average end-to-end delay on the packets. The average delay metric is plotted in Fig. \ref{fig:average_delay_FF} for various values of $\mbox{SNR}$ at a specific value of $N$ and the LOS vector $\mathbf{c}$.
	The plots suggest that the average delay is significantly lower than that of the deadline, especially when $\tau_{NACK}$ is small, owing to the opportunistic nature of CC-HARQ-FF protocol. However, as $\tau_{NACK}$ increases, the average delay is pushed slightly closer to the deadline. Furthermore, to capture the behaviour of deadline violations due to higher $\tau_{NACK}$, in Fig. \ref{fig:average_delay_FF}, we also plot $\eta = \frac{P_{Drop} + P_{Deadline}}{P_{Drop}}$. The plots confirm that when $\tau_{NACK}$ is sufficiently small compared to $\tau_{p} + \tau_{d}$ (see $\tau_{NACK} = 0.05 ~\mu s$ at \mbox{SNR} = 10, 15, 20 dB), the packets that reach the destination arrive within the deadline with an overwhelming probability as $\eta = 1$ at those values.
	
\begin{figure}[h!]
		\centering \includegraphics[scale=0.49]{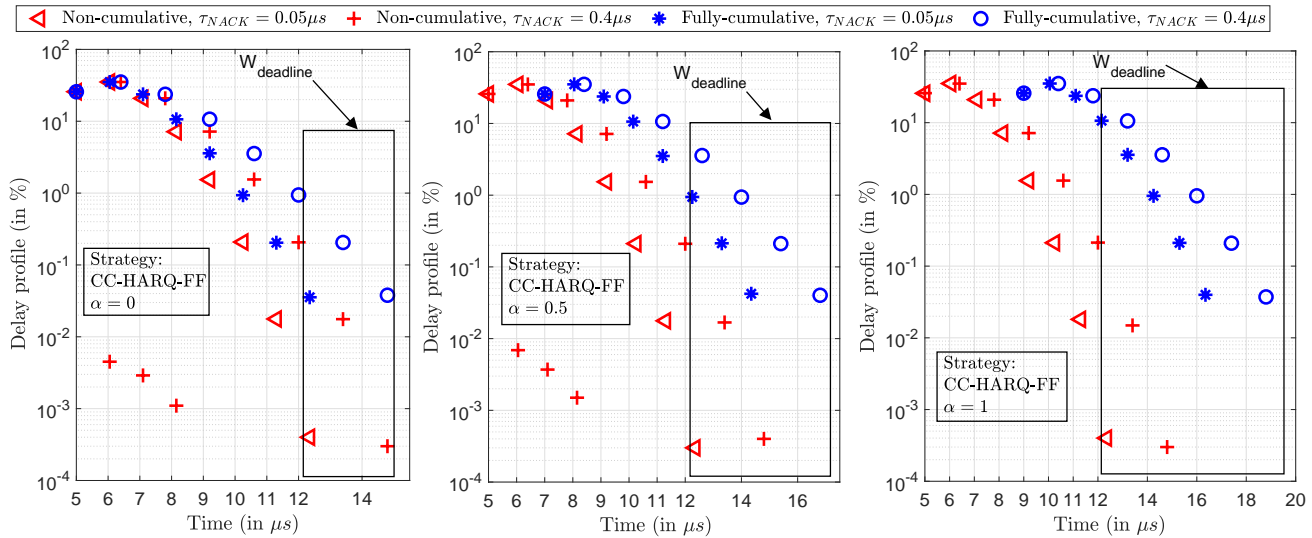}
		\vspace{-0.2cm}
		\centering{\caption{Simulation results on delay profiles (in $\%$) for CC-HARQ-FF based strategies using a $5$-hop network with $\mathbf{c}= [0.1, 0.5, 0.1, 0.3, 0.7]$ and $q_{sum}=12$ at rate $R=1$ and SNR$=5$ dB with $10^{6}$ packets wherein some percentage of packets are dropped either due to outage and some percentage are dropped due to deadline violation, denoted by $W_{\text{deadline}}$ (marked in the rectangle).
				\label{fig:Delay_profile_FF}}}
	\end{figure}

	In the rest of this section, we present other delay metrics for the non-cumulative and fully-cumulative strategies. Similar to Section \ref{sec:Sims_delay_analysis_SF}, we assume that the delay introduced on the packet per hop for each transmission is $T = \tau_{p}+\tau_{d}= 1 ~\mu s$, $\tau_{NACK} \in \{0.05, 0.2, 0.8\}$ in $\mu s$ and $T_{c} = \alpha T$, where $\alpha = 0, 0.5, \mbox{ and }1$. By using these parameters, in Fig. \ref{fig:Delay_profile_FF}, we have shown the delay profiles (in percentage) of both non-cumulative and fully-cumulative strategies. For generating the plots, we considered $N=5$, $q_{sum}=12$, $R=1$, SNR$=5$ dB, and $\mathbf{c} = \{0.1, 0.5, 0.1, 0.3, 0.7\}$ and an ensemble of $10^{6}$ packets. It can be observed that the delay profiles remain unchanged for the non-cumulative strategy irrespective of the value of $\alpha$ because there is no counter present in it. However, it can be observed in a fully-cumulative network that as $\alpha$ increases, the percentage of packets violating the deadline is more. This can be visualized by the width of the rectangle in Fig. \ref{fig:Delay_profile_FF}. 
	Furthermore, when $\alpha=0$ while designing $q_{sum}$, there is a non-zero probability that some packets may reach the destination beyond the deadline; however, it is minimal.
	
	\section{Implementation Aspects and Future Work} 
	\label{sec:discussion}
	
	To implement the non-cumulative strategies (either for slow-fading or fast-fading channels), the nodes would need to share their long-term statistics of their channel to a control center through a backhaul link. Subsequently, the control center would need to solve the optimization problem on optimal ARQ distribution for a given $q_{sum}$, and then distribute the ARQs to the respective nodes. We highlight that this overhead for backhaul coordination is minimal especially when the long-term statistics remain constant for long durations. On the other hand, to implement the fully-cumulative strategy, there is no need for solving the optimization problem since the optimal ARQ distribution is known in closed-form. From a system design's viewpoint, whether to implement non-cumulative or fully-cumulative strategy depends on whether the additional communication-overhead for carrying the counter in the packet is permitted in the protocol. For instance, when $\lceil \mbox{log}_{2} ~q_{sum} \rceil$ is non-negligible compared to the packet size, the system designer may opt for the non-cumulative strategy over the fully-cumulative strategy. In this work, we have presented rigorous analyses of PDP, and have evaluated PDV only as a special case wherein $q_{sum}$ is not estimated accurately. However, it is an interesting problem for future research to analytically characterize PDV, and then minimize the PDP subject to an upper bound on PDV.

\end{document}